\documentclass[11pt,a4paper]{article}

\usepackage[T1]{fontenc}
\usepackage{lmodern}
\usepackage{amsmath,amssymb}
\usepackage{graphicx}
\usepackage{authblk}
\usepackage{geometry}
\usepackage{xcolor}
\usepackage{subcaption}
\usepackage{float}
\usepackage{hyperref}
\usepackage{fancyhdr}
\usepackage{etoolbox}
\usepackage{comment}
\usepackage{subcaption}
\usepackage{tikz}
\usetikzlibrary{positioning, arrows.meta, calc}

\hypersetup{
  colorlinks=true,
  linkcolor=blue,
  citecolor=magenta,
  urlcolor=blue,
}

\fancypagestyle{firstpage}{
  \fancyhf{}

  \fancyfoot[C]{\small Preprint submitted to \textit{Journal Name} \hspace{5mm} }
}
\title{\textbf{Learning the Standard Model Manifold: Bayesian Latent Diffusion for Collider Anomaly Detection}}

\author[1]{Jigar Patel}
\affil[1]{\small Istituto Nazionale di Fisica Nucleare - Sezione di Padova, Italy}
\author[1,2,3,4]{Tommaso Dorigo}
\affil[2]{\small MODE Collaboration}
\affil[3]{\small Lule\aa \, University of Technology, Sweden}
\affil[4]{\small Universal Scientific Education and Research Network (USERN), Italy}

\date{}

\makeatletter
\renewenvironment{abstract}{
  \par\vspace{10pt}
  \hrule height 0.8pt
  \vspace{8pt}
  \begin{center}
  \bfseries \abstractname
  \end{center}
  \vspace{-8pt}
  \itshape
}{
  \vspace{8pt}
  \hrule height 0.8pt
  \vspace{12pt}
}
\makeatother

\newenvironment{keyword}{
  \vspace{-4pt}
  \noindent\textbf{Keywords:}\ }{\vspace{12pt}}

\begin{document}

\maketitle
\thispagestyle{firstpage}

\begin{abstract}
We propose a physics-informed anomaly detection framework for collider data based on a Bayesian latent diffusion model. Our method combines a probabilistic encoder with diffusion dynamics in the latent space, allowing for stable and flexible density estimation while explicitly enforcing physics constraints, such as mass decorrelation and regularization of latent correlations. We train and test the model on simulated LHC jet data and evaluate its performance using seed-averaged ROC curves together with discovery-oriented metrics. Through a series of ablation studies, we show that the diffusion process, Bayesian regularization, and physics-motivated loss terms each contribute in a complementary way: they help stabilize training and improve generalization, even when the gains in peak performance are moderate. Overall, our results emphasize the importance of incorporating both uncertainty estimates and physics consistency when building reliable anomaly detection methods for new Physics searches in high-energy physics.
\end{abstract}



\noindent\rule{\textwidth}{0.4pt}

\vspace{-8pt}
\setcounter{tocdepth}{2}
\renewcommand{\contentsname}{\normalsize \textbf{Contents}}
\tableofcontents
\vspace{10pt}
\hrule height 0.5pt
\vspace{10pt}

\section{Introduction}

The discovery of the Higgs boson—the final missing piece of the Standard Model (SM)—by the ATLAS and CMS collaborations at the LHC in 2012~\cite{ATLAS_Higgs,CMS_Higgs} marked a major milestone in our understanding of fundamental particles and their interactions. Despite its remarkable experimental success, the SM is widely regarded as an effective theory rather than a complete description of nature. Open problems that point toward physics beyond the SM include the origin of neutrino masses~\cite{Fukuda:1998mi}, the structure of the Higgs potential and its self-coupling~\cite{Degrassi_2016,ATLAS:2022jtk,PhysRevLett.133.101801}, the hierarchy problem~\cite{Giudice:2008bi}, the strong CP problem~\cite{Peccei:1977hh}, and the mechanism responsible for the baryon asymmetry of the Universe~\cite{Sakharov:1967dj,Morrissey:2012db}. These unresolved questions provide strong motivation for direct searches for new phenomena at the energy frontier.

The above-mentioned open questions suggest the existence of physics beyond the Standard Model (BSM), motivating extensive experimental and theoretical efforts to search for new phenomena. Despite significant progress, conventional BSM searches at the LHC have not revealed deviations from SM expectations. This has motivated growing interest in model-agnostic strategies that do not depend on specific new-physics hypotheses. Among these, anomaly detection offers a promising path by extending and boosting the power of searches for statistically significant deviations from background predictions, with the potential to uncover unexpected phenomena that might not be captured in traditional signature-driven analyses.

Anomaly detection methods applicable to high-energy physics can be broadly classified into supervised, semi-supervised, and unsupervised approaches~\cite{Chandola2009,Blance2020,Heimel2019,Belis2024MLAnomaly}. Supervised techniques rely on labeled signal and background samples to train classifiers, but their effectiveness is limited by the absence of reliable BSM signal labels and the risk of bias toward specific models~\cite{Guest2018}. Semi-supervised approaches, where models are trained on background-only data and tested against mixed datasets, partially alleviate this issue but still depend on well-defined assumptions about the background distribution~\cite{Cerri:2019variational}. In contrast, unsupervised anomaly detection does not require explicit signal labels or detailed signal hypotheses. By directly modeling SM backgrounds and searching for deviations, these methods offer a more flexible, model-agnostic framework better suited to discovering unexpected signatures at the LHC.

Building on this motivation, unsupervised anomaly detection has been explored extensively through deep generative models such as autoencoders (AEs) and variational autoencoders (VAEs)~\cite{Blance2020,Farina2020,Finke2021,Heimel2019}. These models compress high-dimensional event data into a low-dimensional latent space and attempt to reconstruct them, with the reconstruction error serving as an anomaly score. However, standard AEs often lack uncertainty quantification and can overfit, leading to false positives~\cite{Finke2021,Heimel2019}.

Recently, diffusion probabilistic models (DPMs) have shown impressive generative capabilities across domains such as image synthesis and particle physics~\cite{Ho2020}. In particular, latent diffusion models apply DPMs to compressed representations, improving computational efficiency while retaining expressive power. At the same time, Bayesian neural networks offer principled uncertainty estimates by learning posterior distributions over latent variables~\cite{Blundell2015}.

Recent progress in reliable Bayesian deep learning ~\cite{Fortuin2023BNNPrimer} has further emphasized the importance of calibrated uncertainty in scientific machine learning. Fortuin and collaborators have developed frameworks for uncertainty quantification, model calibration, and Bayesian generative inference that inspire our probabilistic treatment of latent representations. In this work, we extend those principles to the collider-physics domain by combining Bayesian variational encoding with latent diffusion modeling, thereby creating uncertainty-aware representations directly optimized for anomaly detection.

\subsection*{Novelty and Scope of This Work}

Building on the above motivation, our contribution lies in integrating three key advances into a single, physics-aware anomaly-detection framework:
\begin{enumerate}
    \item a \textbf{Bayesian variational encoder} that provides event-wise epistemic and reconstruction uncertainty via stochastic latent representations, rather than uncertainty on physical or model parameters,
    \item a \textbf{latent diffusion process} that models Standard-Model background manifolds in an expressive yet computationally efficient way, and
    \item \textbf{physics-aware regularization} that prevents deformations (“sculpting”) of invariant-mass distributions, which constitute the primary observable in a large class of resonance searches at the LHC. This class of techniques is commonly referred to as \emph{mass decorrelation}\footnote{For an extensive review of the problem and existing approaches, see e.g.~\cite{Larkoski:2017jix,Shimmin2017}.}.
\end{enumerate}

This work presents, to our knowledge, the first integration of Bayesian uncertainty quantification~\cite{Fortuin2023BNNPrimer} with latent-diffusion modeling in an unsupervised collider anomaly-detection framework. Previous diffusion-based studies in high-energy physics~\cite{Mikuni:2022xry} primarily focused on event or detector simulation rather than data-driven discovery. A variety of approaches have addressed mass decorrelation in related contexts—such as adversarial training, distance-correlation penalties, or decorrelated autoencoders used in jet-tagging and anomaly-detection pipelines~\cite{DorigoDeCastro2022,Louppe2017,Shimmin2017,Dolen:2016thinking,Cerri:2019variational}. However, achieving robust mass decorrelation while simultaneously learning a generative, uncertainty-aware latent manifold remains challenging. Our approach contributes to this direction by incorporating physics-motivated decorrelation directly into a Bayesian latent-diffusion architecture.


We validate our approach on benchmark collider datasets with an emphasis on robustness, stability across random seeds, and physics consistency. Rather than focusing solely on mass-peak classification metrics, we assess how individual architectural components influence background modeling, uncertainty calibration, and mass decorrelation through systematic ablation studies. The results demonstrate that Bayesian uncertainty estimation and latent diffusion primarily enhance the stability and interpretability of anomaly scores, while physics-aware regularization suppresses trivial kinematic correlations. Together, these features yield a robust and experimentally relevant framework for model-independent new-physics searches at the LHC.

\section{Related Work and Theoretical Background}

\subsection{Anomaly Detection in High-Energy Physics}

As discussed in introduction section, anomaly detection has emerged as a flexible strategy for searching for new physics beyond traditional, model-specific analyses. Existing approaches differ primarily in their level of supervision, generative modeling capability, and treatment of uncertainty. Here, we summarize a set of representative approaches and highlight their main differences, which helps to place our method in the context of existing work.

The Classification Without Labels (CWoLa) framework introduced a weakly supervised approach that trains classifiers to distinguish mixed samples of signal and background without requiring explicit event-level labels~\cite{CWOLA}. By exploiting differences in signal fractions between datasets, CWoLa can achieve strong discrimination power without explicit background modeling. However, its performance relies on carefully constructed mixed samples and does not provide a generative description of the background, limiting interpretability and control over kinematic correlations such as mass sculpting.

More recently, the Classifying Anomalies Through Outer Density Estimation (CATHODE) method proposed a conditional density-estimation strategy for localized anomaly detection in collider datasets~\cite{Hallin2022}. By learning a background density conditioned on auxiliary variables, CATHODE enables statistically well-defined anomaly tests in specific regions of phase space and represents an important advance over purely discriminative approaches such as CWoLa. At the same time, its reliance on explicit conditioning variables and localized density estimation can limit scalability and robustness when applied to high-dimensional or weakly constrained feature spaces.

The New Physics Learning Machine (NPLM)~\cite{DAgnolo2019} offers an unsupervised approach that models the background likelihood using normalizing flows trained on background-dominated data. This likelihood-based strategy provides a principled anomaly score and explicit density estimation. However, as with other deterministic generative models, controlling correlations between the learned likelihood and kinematic observables such as invariant mass requires additional care, and uncertainty calibration is not intrinsic to the framework.

Generative-model-based approaches, particularly autoencoders (AEs) and variational autoencoders (VAEs)~\cite{Blance2020,Finke2021,Heimel2019}, have been widely explored for unsupervised new-physics searches. These methods compress high-dimensional collider events into low-dimensional latent representations and identify anomalies through reconstruction error. However, most existing AE- and VAE-based approaches are deterministic and lack calibrated uncertainty estimates, which can lead to overfitting or sensitivity to statistical fluctuations in the background.

In contrast, the present work combines uncertainty-aware Bayesian latent encodings with diffusion-based generative modeling within a single framework. Rather than targeting maximal separation power, this design emphasizes statistical robustness, interpretability, and physical consistency. In particular, mass decorrelation is incorporated as an explicit physics-motivated constraint, building on a broad body of previous work that has highlighted the importance of controlling kinematic correlations in anomaly-detection and classification tasks ~\cite{Louppe2017,Shimmin2017,Kasieczka2019JetSubstructure,Nachman2021Review}.

This positioning distinguishes our approach from purely discriminative or deterministic anomaly-detection strategies by prioritizing stable behavior across random seeds, controlled correlations with invariant mass, and reproducible uncertainty estimates. 

\subsection{Bayesian Neural Networks for Uncertainty Quantification}

Bayesian neural networks (BNNs) provide a principled framework for modeling uncertainty by defining probability distributions over network weights or latent variables~\cite{Blundell2015}. Techniques such as variational inference~\cite{Blundell2015}, Monte Carlo dropout~\cite{Gal2016}, and approximate Bayesian inference in deep generative models~\cite{pmlr-v32-rezende14} have been successfully applied to encode event-level uncertainty in collider physics and other scientific domains.

Recent progress in reliable Bayesian deep learning~\cite{Fortuin2023BNNPrimer} has highlighted the importance of calibrated uncertainty estimates and robust posterior inference, particularly in high-stakes scientific applications. These developments motivate the probabilistic design of our encoder, which models uncertainty at the level of latent representations rather than physical parameters. When combined with generative modeling, this uncertainty-aware formulation primarily contributes to stabilizing training across stochastic realizations and supporting more interpretable anomaly scores.

Consistent with these expectations, our ablation studies show that Bayesian latent modeling primarily improves stability across random seeds and supports more reliable anomaly scoring, rather than yielding large gains in peak classification performance.

\subsection{Diffusion Probabilistic Models and Latent Diffusion}

Denoising Diffusion Probabilistic Models (DDPMs) define a Markovian generative process consisting of a forward noising procedure and a learned reverse denoising dynamics, enabling expressive and stable generative modeling~\cite{Ho2020}. In high-energy physics, diffusion-based approaches have recently been applied to event generation and detector simulation tasks~\cite{Mikuni:2022xry}.

Latent Diffusion Models (LDMs)~\cite{Rombach2022} improve computational efficiency by performing diffusion in a compressed latent space rather than directly in the high-dimensional data space. This strategy is particularly well suited to collider datasets, where raw-space diffusion can be prohibitively expensive and sensitive to noise. In the present work, latent diffusion acts as a generative regularizer on the learned background manifold, smoothing latent representations and reducing sensitivity to statistical fluctuations. This role complements the Bayesian encoder by promoting stability and consistency rather than solely enhancing reconstruction fidelity.

In practice, we find that the latent diffusion mainly smooths the learned background manifold and makes the model less sensitive to statistical fluctuations, which becomes clear when we remove the diffusion component in ablation studies.
\section{Methodology}
\label{sec:methodology}

The proposed anomaly-detection framework unifies probabilistic representation learning with generative modeling in a physically constrained latent space. Rather than targeting maximal classification performance, the design emphasizes robustness, stability across stochastic training realizations, and physics consistency. The framework consists of three key components: a Bayesian variational encoder, a latent-space diffusion model, and a reconstruction decoder. Physics-aware loss terms explicitly constrain the learned representations to preserve essential kinematic and substructure properties throughout training.

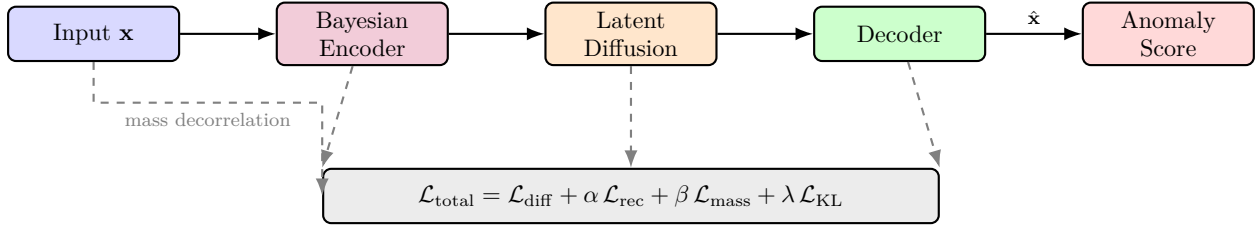
\begin{figure}[htbp]
\centering
\begin{tikzpicture}[scale=0.9, transform shape,
    node distance=1.4cm,
    every node/.style={font=\small},
    box/.style={
        draw, rounded corners=3pt,
        minimum width=2.5cm, minimum height=0.8cm,
        align=center, thick
    },
    arrow/.style={->, thick, >=Latex},
    darrow/.style={->, thick, >=Latex, dashed, gray}
]

\node[box, fill=blue!15]    (input)    {Input $\mathbf{x}$};
\node[box, fill=purple!20,  right=of input]    (encoder)  {Bayesian\\Encoder};
\node[box, fill=orange!20,  right=of encoder]  (diffusion){Latent\\Diffusion};
\node[box, fill=green!20,   right=of diffusion](decoder)  {Decoder};
\node[box, fill=red!15,     right=of decoder]  (score)    {Anomaly\\Score};

\draw[arrow] (input)    -- (encoder);
\draw[arrow] (encoder)  -- (diffusion);
\draw[arrow] (diffusion)-- (decoder);
\draw[arrow] (decoder)  -- node[above, font=\scriptsize]
             {$\hat{\mathbf{x}}$} (score);

\node[box, fill=gray!15,
      below=1.5cm of diffusion,
      minimum width=9cm] (loss)
    {$\mathcal{L}_{\mathrm{total}} =
      \mathcal{L}_{\mathrm{diff}}
      + \alpha\,\mathcal{L}_{\mathrm{rec}}
      + \beta\,\mathcal{L}_{\mathrm{mass}}
      + \lambda\,\mathcal{L}_{\mathrm{KL}}$};

\draw[darrow] (encoder)   -- (loss.north west);
\draw[darrow] (diffusion) -- (loss.north);
\draw[darrow] (decoder)   -- (loss.north east);
\draw[darrow] (input.south) -- ++(0,-0.6) -|
    node[pos=0.25, below, font=\scriptsize]{mass decorrelation}
    (loss.west);

\end{tikzpicture}
\caption{
    Overview of the Bayesian latent diffusion framework. Input events 
    are encoded into stochastic latent representations, refined by 
    latent diffusion, and reconstructed by the decoder. The anomaly 
    score is derived from the uncertainty-normalized reconstruction 
    error, and training is governed by the combined physics-aware 
    objective $\mathcal{L}_{\mathrm{total}}$.
}
\label{fig:model_diagram}
\end{figure}

As shown in Figure~\ref{fig:model_diagram}, collider events are first encoded into stochastic latent representations by the Bayesian encoder, which captures event-level uncertainty. The latent diffusion model then acts as a generative regularizer, smoothing the learned background manifold through iterative noising and denoising steps. The decoder reconstructs high-dimensional observables from the denoised latent variables. Training is guided by reconstruction, diffusion, and physics-aware regularization terms, while anomaly scores combine reconstruction error with predictive uncertainty to identify deviations from SM expectations.

\subsection{Bayesian Variational Encoding of Collider Events}

Given an input event $x \in \mathbb{R}^D$, a Bayesian variational encoder maps $x$ into a latent distribution $q_\phi(z|x)$,where $z \in \mathbb{R}^d$ denotes a compact stochastic embedding and $\phi$ the encoder parameters. Prior to training, all input features are standardized to zero mean and unit variance using statistics computed on background-only samples, ensuring that the reconstruction loss is well-defined and balanced across observables. The encoder outputs a mean vector and log-variance, thereby modeling epistemic uncertainty in the representation.

Training follows the standard variational-inference objective by maximizing the evidence lower bound (ELBO):

\begin{equation}
\mathcal{J}_{\mathrm{ELBO}} = \mathbb{E}_{q_\phi(z|x)} \left[ \log p_\theta(x|z) \right] - \mathrm{KL}\big(q_\phi(z|x) \| p(z)\big)
\end{equation}

\begin{equation}
\mathcal{L}_{\mathrm{VAE}} = - \mathcal{J}_{\mathrm{ELBO}} 
\end{equation}

where $p(z) = \mathcal{N}(0, I)$ is a standard Gaussian prior. We adopt the convention that objectives $\mathcal{J}_{\mathrm{ELBO}}$ are maximized, while losses $\mathcal{L}_{\mathrm{VAE}}$ are minimized. The KL term regularizes the latent space and propagates into the diffusion stage.

In practice, this Bayesian formulation primarily stabilizes latent representations across random seeds and provides calibrated uncertainty estimates for anomaly scoring, rather than directly maximizing reconstruction fidelity.

\subsection{Latent Diffusion Modeling with KL Regularization}

To model the latent distribution generatively, we employ a denoising diffusion probabilistic model (DDPM) on $z \sim q_\phi(z|x)$. The forward process gradually adds Gaussian noise over $T$ steps:

\begin{equation}
q(z_t | z_{t-1}) = \mathcal{N}(z_t; \sqrt{1 - \beta_t}\, z_{t-1}, \beta_t \mathbf{I}),
\end{equation}

A learnable reverse process parameterized by $\epsilon_\theta(z_t, t)$ predicts the added noise $\epsilon$ at each step, using time embedding and conditioning on $z_t$.

The diffusion loss combines the noise-prediction objective with KL regularization from the encoder:

\begin{equation}
\mathcal{L}_{\mathrm{diff}} = \mathbb{E}_{z_0, \epsilon, t} \!\left[\! \| \epsilon - \epsilon_\theta(z_t, t) \|^2 \!\right]
 + \lambda_{\mathrm{KL}}\, \mathrm{KL}\big(q_\phi(z|x)\|p(z)\big),
\end{equation}

where $\lambda_{\mathrm{KL}}$ controls the regularization strength.

Including the KL term in the diffusion objective ensures consistency between the stochastic encoder posterior and the generative prior assumed by the diffusion process, thereby preventing latent-space drift during training.

\subsection{Decoder and Reconstruction Loss}

Following denoising, a deterministic decoder $p_\theta(x|z)$ reconstructs event features from the denoised latent vector. The reconstruction objective is the mean-squared error (MSE) evaluated over standardized input features:

\begin{equation}
\mathcal{L}_{\mathrm{rec}} = \| x - \hat{x} \|^2, \quad \hat{x} = p_\theta(z).
\end{equation}

The decoder is implemented as a feed-forward network mapping denoised latent vectors back to observable space, providing a reconstruction signal suitable for anomaly detection.

\subsection{Physics-Aware Regularization}

To maintain physical interpretability, the model is trained with additional physics-motivated regularization terms, each contributing explicitly to the total training objective with tunable weights:

\begin{itemize}
    \item \textbf{Mass Decorrelation Loss:} Penalizes correlations between the anomaly score and a designated kinematic control variable whose distribution is intended to be preserved, such as the invariant mass in resonance-search applications:
    
    \begin{equation}
    \mathcal{L}_{\mathrm{mass}} = \mathrm{MSE}(m_{\mathrm{reco}}, m_{\mathrm{input}}).
    \end{equation}

    This term suppresses mass-dependent structure in the anomaly score by discouraging direct propagation of invariant-mass information through the reconstruction pathway. We emphasize that it acts as a soft regularizer rather than enforcing exact agreement between mass distributions. While distribution-level divergences such as Jensen–Shannon or Wasserstein metrics could be used to explicitly match spectra, the present formulation is validated a posteriori through explicit mass–score correlation measurements and ablation studies presented in Sec.~\ref{sec:mass_decorrelation}.

    \item \textbf{Latent Diffusion and KL Loss:} Combines the diffusion noise-prediction term with KL divergence from the encoder to enforce regularization and maintain consistent posteriors.

    \item \textbf{Reconstruction Loss:} Ensures high-fidelity recovery of input observables from denoised latent embeddings.

    \item \textbf{Substructure Fidelity Metrics:} Correlations of jet substructure observables $(\tau_1,\tau_2,\tau_3)$ are monitored during validation to confirm physics fidelity, though not directly included in the training loss.
\end{itemize}

\subsection{Overall Objective}

The full training objective integrates all components:

\begin{equation}
\mathcal{L}_{\mathrm{total}} = \mathcal{L}_{\mathrm{diff}} + \alpha\, \mathcal{L}_{\mathrm{rec}} + \beta\, \mathcal{L}_{\mathrm{mass}},
\end{equation}

where $\alpha$ and $\beta$ are scalar hyperparameters. This objective balances expressivity and regularization, encouraging a structured latent space that captures dominant background features while remaining insensitive to trivial kinematic correlations.

\subsection{Anomaly Scoring}

Event-level anomaly scores combine normalized reconstruction error and predictive uncertainty:
\[
\mathrm{Score}(x) = \frac{\| x - \hat{x} \|^2}{\sigma_{\hat{x}}},
\]
where $\sigma_{\hat{x}}$ denotes the reconstruction uncertainty estimated from multiple stochastic forward passes through the Bayesian encoder, reflecting epistemic uncertainty in the latent representation.

This uncertainty-aware normalization suppresses spurious contributions from poorly constrained regions of latent space while emphasizing confident deviations from the learned SM background. As demonstrated in the results section, this formulation primarily improves the stability and interpretability of anomaly scores rather than maximizing raw separation power.

\subsection{Physics-Motivated Ablation and Architecture Study}

We perform a structured set of controlled ablation and sensitivity studies to systematically assess the physical and architectural contributions of each model component. These experiments are designed not to optimize headline performance metrics, but to isolate cause--effect relationships between physics constraints, representational choices, and observed anomaly-detection behavior. All studies are conducted using identical datasets, training procedures, and evaluation metrics, with multiple random seeds to ensure statistical robustness.

The experimental program is organized into three complementary categories: physics-motivated ablations, representational capacity studies, and generative modeling sensitivity analyses.

\paragraph{Baseline configuration:}
The baseline model incorporates all proposed elements: a Bayesian variational encoder with KL regularization, a latent-space diffusion model, explicit mass decorrelation regularization, and a fixed latent dimensionality. This configuration serves as the reference point for all subsequent comparisons and defines the fully physics-aware anomaly-detection framework studied in this work.

\paragraph{Physics-motivated ablations:}
To isolate the role of individual physical assumptions, we consider the following targeted modifications:
\begin{itemize}
    \item \textbf{No mass decorrelation:} The mass decorrelation term is disabled, allowing the anomaly score to correlate freely with the reconstructed invariant mass. This configuration probes the extent to which apparent performance gains can arise from exploiting kinematic mass correlations rather than genuine substructure anomalies, and tests the necessity of explicit de-sculpting constraints for physically meaningful anomaly detection.

    \item \textbf{No KL regularization:} The KL divergence term in the Bayesian encoder is removed, resulting in deterministic latent embeddings. This ablation evaluates the role of uncertainty-aware representations in stabilizing training and anomaly scoring across stochastic realizations.

    \item \textbf{No latent diffusion:} The diffusion module is removed, and the decoder operates directly on the encoder output. This configuration tests whether generative smoothing of the latent space is required to robustly model the SM background manifold.
\end{itemize}

\paragraph{Latent representational capacity.}
To study the impact of latent-space dimensionality on anomaly detection behavior, we perform a sweep over multiple latent dimensions ($d = 4, 8, 16, 32$). This analysis probes the trade-off between information compression, representational expressivity, and over-regularization, and helps distinguish performance variations driven by latent capacity from those arising from physics-aware regularization.

\paragraph{Diffusion process sensitivity.}
The influence of the diffusion time horizon is examined by varying the number of diffusion steps ($T = 200, 500, 1000$). This study assesses whether diffusion primarily enhances generative fidelity or instead acts as a regularizing mechanism that smooths the learned background manifold and reduces sensitivity to statistical fluctuations.

\medskip
Importantly, none of these variations introduce additional supervision or modify the definition of anomalies. Instead, they serve as controlled diagnostics to validate the physical necessity, robustness, and scalability of each component in the proposed framework. As shown in Section~\ref{sec:results}, these studies demonstrate that improvements in stability, interpretability, and physics consistency are driven by the inclusion of Bayesian uncertainty modeling, latent diffusion regularization, and mass decorrelation, rather than by increased model complexity alone.

\section{Data and Training}
\label{sec:data_training}

\subsection{Datasets}
\label{subsec:datasets}

The experiments are conducted using the LHCOlympics 2020 dataset~\cite{LHCOlympics2020}, a public benchmark designed to evaluate model-agnostic anomaly-detection strategies under realistic collider conditions. The dataset consists of simulated proton--proton dijet events and is explicitly constructed to probe robustness against modeling uncertainties and unknown signal hypotheses. We adopt the standard preprocessing released with the HEPxML 2023 unsupervised-learning benchmark~\cite{HEPxML2023}, ensuring direct comparability with existing methods and full analysis reproducibility.

The use of different event generators for training and validation intentionally introduces a controlled domain shift, reflecting realistic theoretical uncertainties in Standard Model background modeling at the LHC.

\begin{itemize}
    \item QCD background generated with \textsc{Herwig}~\cite{Herwig7}, used exclusively for training,
    \item QCD background generated with \textsc{Pythia8}~\cite{Pythia8}, used for validation,
    \item A beyond-the-Standard-Model signal process, $W' \rightarrow jj$, generated with \textsc{Pythia8}~\cite{Pythia8}.
\end{itemize}

Each event is represented by 14 high-level observables derived from the two leading jets:
\[
(p_x, p_y, p_z, m, \tau_1, \tau_2, \tau_3)_{\text{jet 1, jet 2}},
\]
where $m$ denotes the jet invariant mass and $(\tau_1, \tau_2, \tau_3)$ are $N$-subjettiness observables capturing jet substructure information.

These high-level observables capture both kinematic information and jet substructure, providing a compact yet physically meaningful representation commonly used in resonance and anomaly searches at the LHC~\cite{Kasieczka2019JetSubstructure,Nachman2021Review}.

\subsection{Preprocessing and Normalization}
\label{sec:preprocessing}

All features are standardized using the mean and standard deviation  computed solely from the QCD (\textsc{Herwig}) training sample. The same  normalization parameters are subsequently applied to validation and signal  events. This background-only normalization avoids information leakage and ensures that anomaly scores are defined relative to a background reference, consistent with unsupervised discovery strategies at the LHC~\cite{AnomalyReview2021,LHCOlympics2020}.

This setup mirrors realistic discovery scenarios, where anomaly-detection models must learn from imperfect background simulations without access to labeled signal samples. No explicit mass-window selections, signal-dependent cuts, or class-conditional preprocessing steps are applied. This preserves the model's agnosticism with respect to the underlying signal hypothesis and prevents artificial enhancement of mass-dependent features.

\begin{figure*}[htbp]
  \centering
  \includegraphics[width=0.87\textwidth]{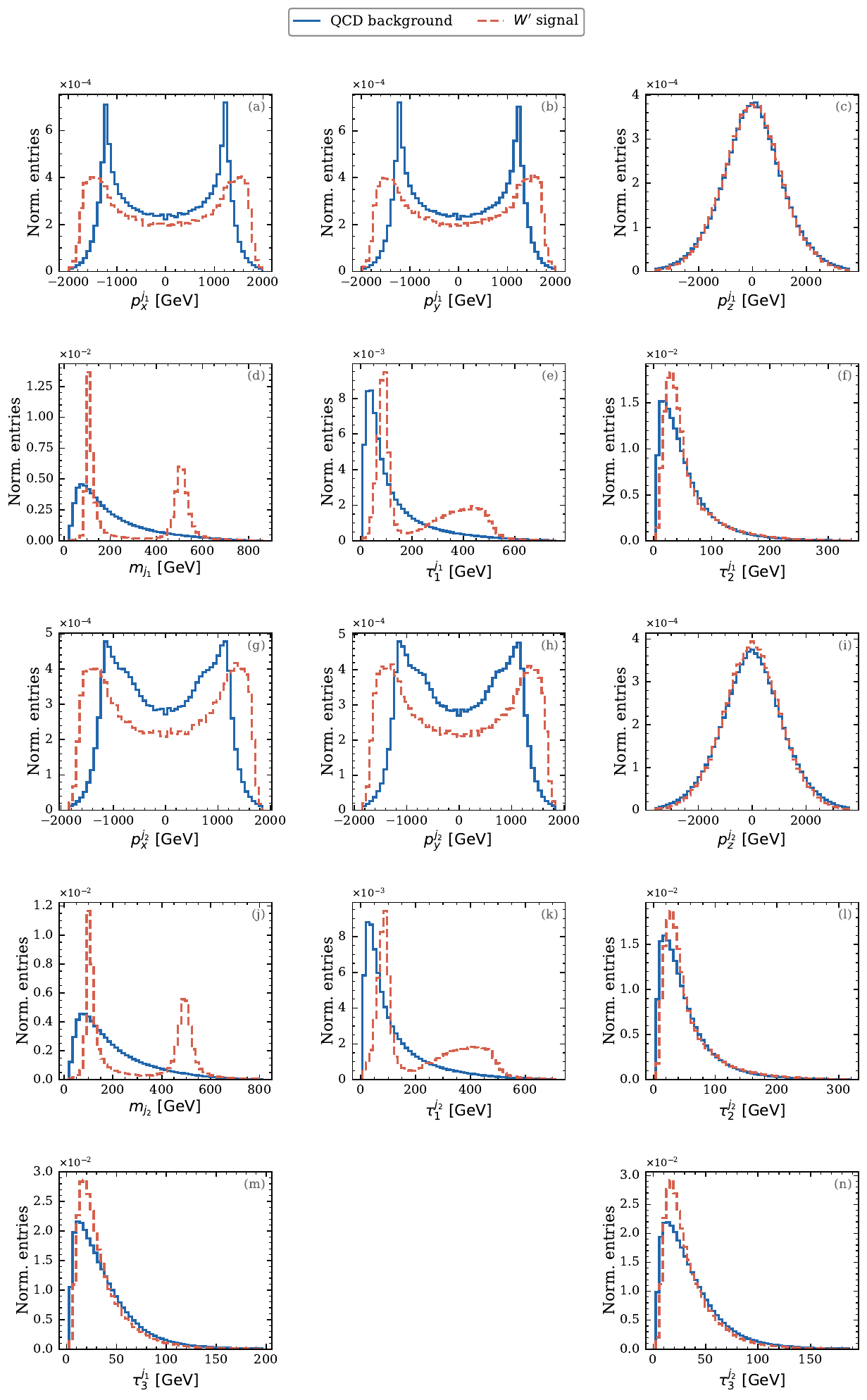}
  \caption{
    Normalized distributions of the 14 input features for the QCD background (blue solid) and the $W^\prime$ signal (red dashed), prior to any selection on the anomaly score. Rows~1--2 correspond to the leading jet $j_1$; rows~3--4 to the subleading jet $j_2$; and the final row presents $\tau_3$ for both jets. All distributions are normalized to unit area.
  }
  \label{fig:dataset_overview}
\end{figure*}

Figure~\ref{fig:dataset_overview} shows the normalized distributions of the 14 input features for the QCD background and the $W^\prime$ signal  prior to any selection. Several physically well-motivated differences are immediately visible. The transverse momentum components $p_x^{j_1}$ and $p_y^{j_1}$ (panels~(a) and~(b)) exhibit the expected bimodal structure arising from the approximately back-to-back dijet topology enforced by transverse momentum conservation. While this feature is present in both samples, the QCD background populates the two lobes more broadly, whereas the $W^\prime$ signal is more localized. The longitudinal momentum $p_z$ (panels~(c) and~(i)) is approximately Gaussian and shows no significant shape difference between signal and background, consistent with the symmetric rapidity acceptance of the detector.

The jet invariant masses $m_{j_1}$ and $m_{j_2}$ (panels~(d) and~(j)) provide the strongest discrimination. The QCD background peaks at low mass and falls steeply, as expected for light-quark and gluon jets whose mass is generated by soft and collinear radiation. In contrast, the $W^\prime$ signal develops a pronounced secondary peak around $m \approx 80$--$100~\mathrm{GeV}$, corresponding to hadronic decays of on-shell $W$ bosons, in addition to a low-mass component from the harder of the two decay jets. This two-peak structure constitutes the dominant handle for separating signal from background and directly motivates the mass-decorrelation constraint introduced in Section ~\ref{sec:methodology}.

The N-subjettiness observables $\tau_1^{j_1}$ and $\tau_1^{j_2}$ (panels~(e) and~(k)) further highlight the difference in jet substructure. QCD jets are concentrated at low $\tau_1$, consistent with collimated single-prong radiation, whereas the $W^\prime$ signal exhibits a broader population extending to larger values, reflecting the two-prong substructure of boosted hadronic $W$ decays. A similar but less pronounced behaviour is observed for $\tau_2$ (panels~(f) and~(l)). The $\tau_3$ distributions (panels~(m) and~(n)) are largely compatible between signal and background, with only small differences at high values, indicating that three-prong substructure does not play a major role in this topology.

The jet mass panels~(d) and~(j) exhibit the most pronounced separation, with the $W^\prime$ signal developing a secondary peak near $m \approx 80$--$100~\mathrm{GeV}$ from hadronic on-shell $W$ decays, while the QCD background retains its characteristic steeply falling spectrum. The $\tau_1$ and $\tau_2$ observables further reveal the presence of two-prong substructure in the signal that is largely absent in the QCD background, providing complementary discrimination beyond the mass alone.

\subsection{Training Procedure}
\label{sec:training_procedure}

The model is trained in a fully unsupervised manner using only QCD (\textsc{HERWIG}) events. Optimization is performed using the ADAM optimizer~\cite{KingmaAdam}, with fixed batch size and component-specific learning rates for the encoder, decoder, and diffusion denoiser. Training proceeds for a fixed number of epochs, with convergence monitored using validation loss and physics-motivated diagnostics.

\begin{table}[htbp]
\centering
\caption{Baseline training hyperparameters for the full model.}
\label{tab:training_hyperparams}
\begin{tabular}{l c}
\hline
\textbf{Parameter} & \textbf{Value} \\
\hline
Latent dimension $d$ & 16 \\
Diffusion steps $T$ & 500 \\
Training epochs & 50 \\
Batch size & 128 \\
Reconstruction loss weight $\alpha$ & 1.0 \\
Diffusion loss weight $w_{\mathrm{diff}}$ & 1.3 \\
Mass decorrelation weight $w_{\mathrm{mass}}$ & 85 (cosine ramp) \\
KL divergence weight $w_{\mathrm{KL}}$ & 0.10 (warm-up) \\
KL warm-up steps & 3000 \\
Optimizer & Adam \\
Learning rate & $3\times10^{-4}$ \\
\hline
\end{tabular}
\end{table}

The baseline training configuration and loss weights are summarized in Table~\ref{tab:training_hyperparams}. 
The relative weights of the reconstruction, diffusion, mass-decorrelation, and KL regularization terms were tuned through a limited set of exploratory runs around physically motivated baseline values (see Appendix~\ref{app:training_loss_all_components} for the corresponding training loss plots). 
The tuning procedure prioritized stable convergence, suppression of spurious kinematic correlations, and reproducibility across random seeds rather than maximizing peak anomaly detection performance. In particular, the mass-decorrelation and KL terms were increased until
residual mass–score correlations were visibly reduced, while preserving stable optimization of
the reconstruction and diffusion objectives. We observe qualitatively stable behavior across a
broad range of nearby hyperparameter choices, indicating that the reported results are not
sensitive to fine tuning.

To ensure statistical robustness, each configuration is trained using multiple random seeds 
\[
\{7,\;42,\;123,\;555,\;1251,\;2024\}.
\] The reported performance metrics correspond to the seed-averaged mean, and the shaded regions in the figures represent the envelope of the resulting run-to-run variation. Such multi-seed evaluation is particularly important for unsupervised and generative models, where stochastic optimization and sampling effects can induce significant variability. This strategy enables a clear separation between genuine architectural or loss-driven improvements and fluctuations arising from training noise~\cite{Reimers2017Reporting}.

The baseline configuration employs a latent dimension of $d=16$ and a diffusion process with $T=1000$ steps. Systematic variations around this baseline are explored through targeted ablation studies described in Section~\ref{sec:ablation_strategy}. Unless explicitly stated otherwise, all hyperparameters listed in Table~\ref{tab:training_hyperparams} are held fixed to isolate the effect of individual architectural or physics-guided components.

\subsection{Physics-Guided Training Monitoring}
\label{sec:training_monitoring}

Beyond monitoring the total training loss, individual loss components corresponding to reconstruction, diffusion, KL regularization, and mass decorrelation are tracked throughout training. This allows verification that no single term dominates optimization or collapses prematurely.

In addition, physics fidelity is assessed by evaluating Pearson correlations between reconstructed and true jet substructure observables $(\tau_1, \tau_2, \tau_3)$ on validation data. Maintaining stable correlations over training ensures that the latent representation preserves physically meaningful information rather than exploiting trivial kinematic correlations or training instabilities, a known failure mode of purely data-driven anomaly detectors~\cite{PhysicsAwareAD,Kasieczka2021MLReview}.

\begin{figure}[htbp]
    \centering
    \includegraphics[width=0.8\textwidth]{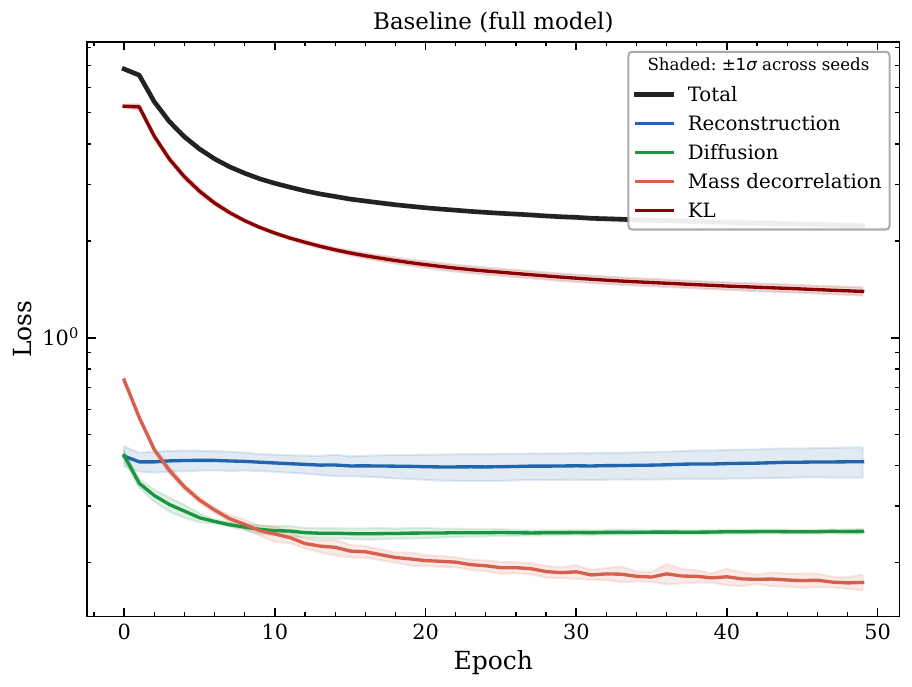}
    \caption{
    Seed-averaged evolution of the total training loss and its individual components for the
    baseline model. Shaded bands indicate the standard deviation across six independent random
    seeds, demonstrating stable and well-balanced optimization.
    }
    \label{fig:training_losses}
\end{figure}

Figure~\ref{fig:training_losses} shows the seed-averaged evolution of the total loss and its
components for the baseline configuration. All loss terms decrease smoothly and stabilize at
late epochs, indicating a well-behaved optimization process. The comparable magnitudes of the
individual contributions demonstrate that physics-motivated constraints act as regularizers
rather than overpowering the reconstruction objective.

\begin{figure}[htbp]
    \centering
    \includegraphics[width=0.75\textwidth]{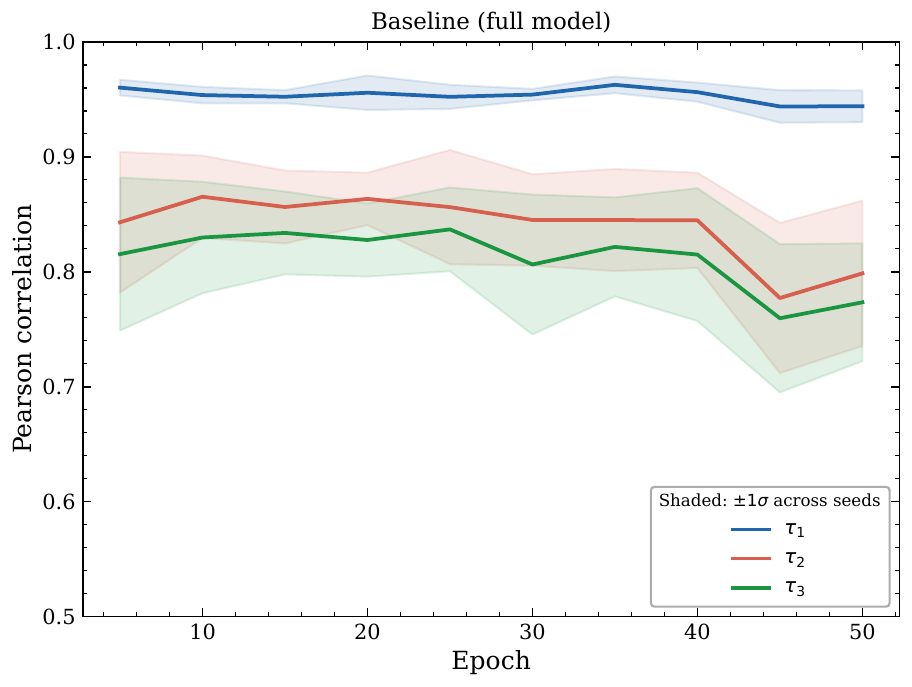}
    \caption{
    Seed-averaged Pearson correlation coefficients between the anomaly score and the jet
    substructure observables $\tau_1$, $\tau_2$, and $\tau_3$ during training. The smooth
    convergence and narrow uncertainty bands indicate stable physics-aware learning.
    }
    \label{fig:tau_correlations}
\end{figure}

In addition to loss convergence, physics fidelity is assessed by monitoring correlations between
the learned anomaly score and jet substructure observables $(\tau_1, \tau_2, \tau_3)$ on validation
data. Figure~\ref{fig:tau_correlations} shows the evolution of the Pearson correlation coefficients
for these observables throughout training. The correlations evolve smoothly and converge to stable
values, indicating that the latent representation preserves physically meaningful information
rather than exploiting unphysical shortcuts.

All training diagnostics are evaluated independently for each random seed and subsequently averaged
at fixed epoch. Shaded bands indicate the standard deviation across seeds, demonstrating that the
training dynamics are stable and reproducible despite the stochastic nature of unsupervised
optimization. Additional per-seed and ablation-specific training curves are provided in
Appendix~\ref{app:training_diagnostics}.

\subsection{Evaluation Protocol}
\label{sec:evaluation_protocol}

The evaluation protocol is designed to probe not only anomaly–background separation, but also robustness to modeling uncertainties and compatibility with data-driven background estimation strategies.

The area under the receiver operating characteristic curve (AUC) and the average precision (AP) quantify the ranking quality of the anomaly score across thresholds, with AP emphasizing performance in the high-score tail most relevant for rare-signal searches~\cite{Davis2006PR,AUCAPHEP}. To connect anomaly detection with experimental discovery reach, the effective significance $Z_{\mathrm{eff}}$ is computed at optimized operating points, following standard approximations for background-dominated searches~\cite{CowanAsimov,LHCOlympics2020}.

For models incorporating Bayesian uncertainty estimation, an uncertainty-filtered anomaly score is constructed by suppressing events with large predictive uncertainty. This procedure mitigates spurious anomalies arising from poorly constrained regions of phase space and improves robustness in fully unsupervised settings~\cite{Kendall2017Uncertainty,Yong2022BayesianAE}.

Finally, the Pearson correlation between the anomaly score and the reconstructed invariant mass is measured to quantify mass sculpting effects. Low mass–score correlation is essential for preserving data-driven background estimation strategies commonly employed in resonance searches~\cite{CWOLA,Louppe2017}.

\section{Results}
\label{sec:results}

\subsection{Baseline Model Performance}

\begin{figure}[htbp]
    \centering
    \includegraphics[width=0.65\textwidth]{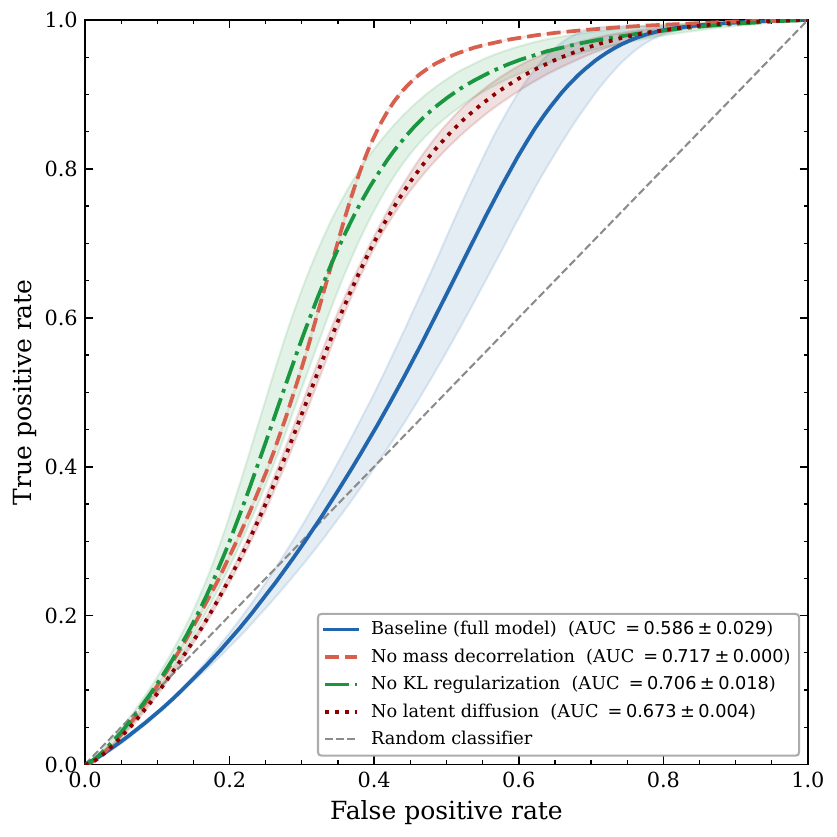}
    \caption{
    Seed-averaged ROC curve for the baseline model. The solid line shows the mean ROC across six independent random seeds, while the shaded band represents the corresponding standard deviation. The dashed line indicates the performance of a random classifier.
    }
    \label{fig:baseline_roc}
\end{figure}

The baseline model is evaluated using the full physics-informed training configuration, which combines mass decorrelation, Bayesian regularization, and latent diffusion. The receiver operating characteristic (ROC) curve averaged over six independent random seeds is shown in Fig.~\ref{fig:baseline_roc}. The solid line represents the mean true positive rate at fixed false positive rate, while the shaded band indicates the seed-to-seed variation.

The model achieves a mean area under the ROC curve of $\mathrm{AUC} = 0.59 \pm 0.03$, demonstrating non-trivial discrimination between QCD background and signal-like events. The relatively narrow uncertainty band across the full ROC range indicates stable classifier behavior and limited sensitivity to stochastic training effects. This robustness is essential for reliable downstream significance estimates and physics interpretation.

\begin{figure}[htbp]
    \centering
    \includegraphics[width=0.65\textwidth]{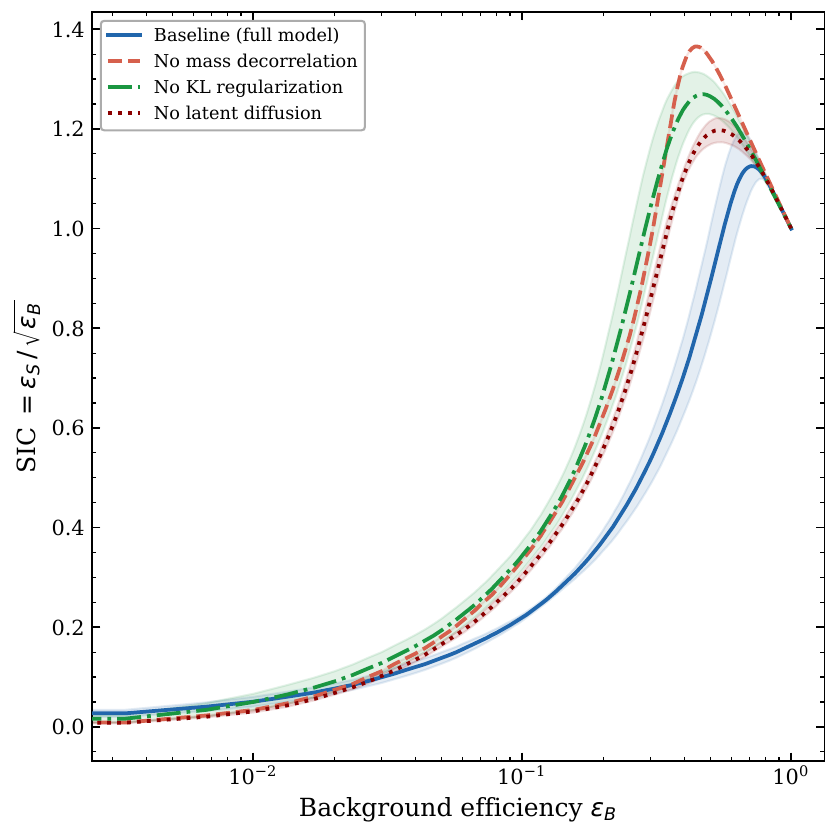}
    \caption{
    Seed-averaged Signal Improvement Characteristic (SIC) curves for the baseline and ablation models. The SIC is defined as $\epsilon_S \sqrt{\epsilon_B}$ and quantifies the improvement
    in discovery significance after a selection on the anomaly score. Solid lines show the mean
    over six independent random seeds, while the shaded bands indicate the corresponding standard
    deviation.
    }
    \label{fig:baseline_sic}
\end{figure}

The expected discovery sensitivity is further quantified using the Signal Improvement Characteristic (SIC), defined as $\epsilon_S / \sqrt{\epsilon_B}$, and the corresponding effective significance $Z_{\rm eff}$. The seed-averaged SIC curves for the baseline and ablation models are shown in Fig.~\ref{fig:baseline_sic}. The baseline configuration exhibits a clear maximum in the low-background regime, corresponding to the anomaly-score threshold that optimizes the expected significance. The small seed-to-seed variation around this maximum confirms the robustness of the classifier and enables a stable choice of operating point for downstream analyses.

\begin{figure}[htbp]
    \centering
    \includegraphics[width=0.65\textwidth]{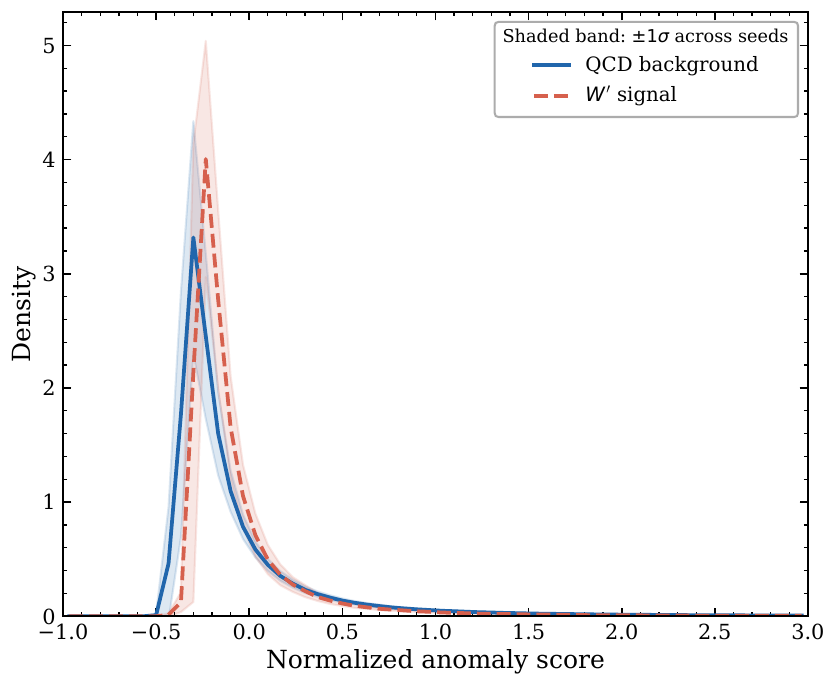}
    \caption{
    Seed-averaged normalized anomaly score distributions for QCD (background) and W$^\prime$
    (signal) events. Solid lines show the mean distributions across six seeds, and shaded bands
    indicate the standard deviation.
    }
    \label{fig:baseline_score_dist}
\end{figure}

Additional insight into the discriminative behavior is provided by the anomaly-score distributions shown in Fig.~\ref{fig:baseline_score_dist}. Signal events are preferentially assigned higher scores, while QCD background events populate the low-score region. This separation is consistent across random seeds and complements the ROC and SIC analyses by directly visualizing the enrichment of signal-like events at high anomaly scores.

A key requirement for resonance searches is that the anomaly score be minimally correlated with the reconstructed invariant mass. For the baseline model, the seed-averaged Pearson correlation coefficient is $\rho(m,\mathrm{score}) = -0.10 \pm 0.03$, indicating a negligible dependence on mass. This demonstrates that the classifier does not exploit trivial kinematic features and that the observed discrimination arises from jet substructure and latent-space geometry. Such decorrelation ensures that selections based on the anomaly score do not sculpt the background mass distribution, preserving the validity of sideband-based background estimation strategies.

\subsubsection{Feature Distributions Before and After Anomaly Score Selection}

\begin{figure}[p]
  \centering
  \includegraphics[width=0.80\textwidth]{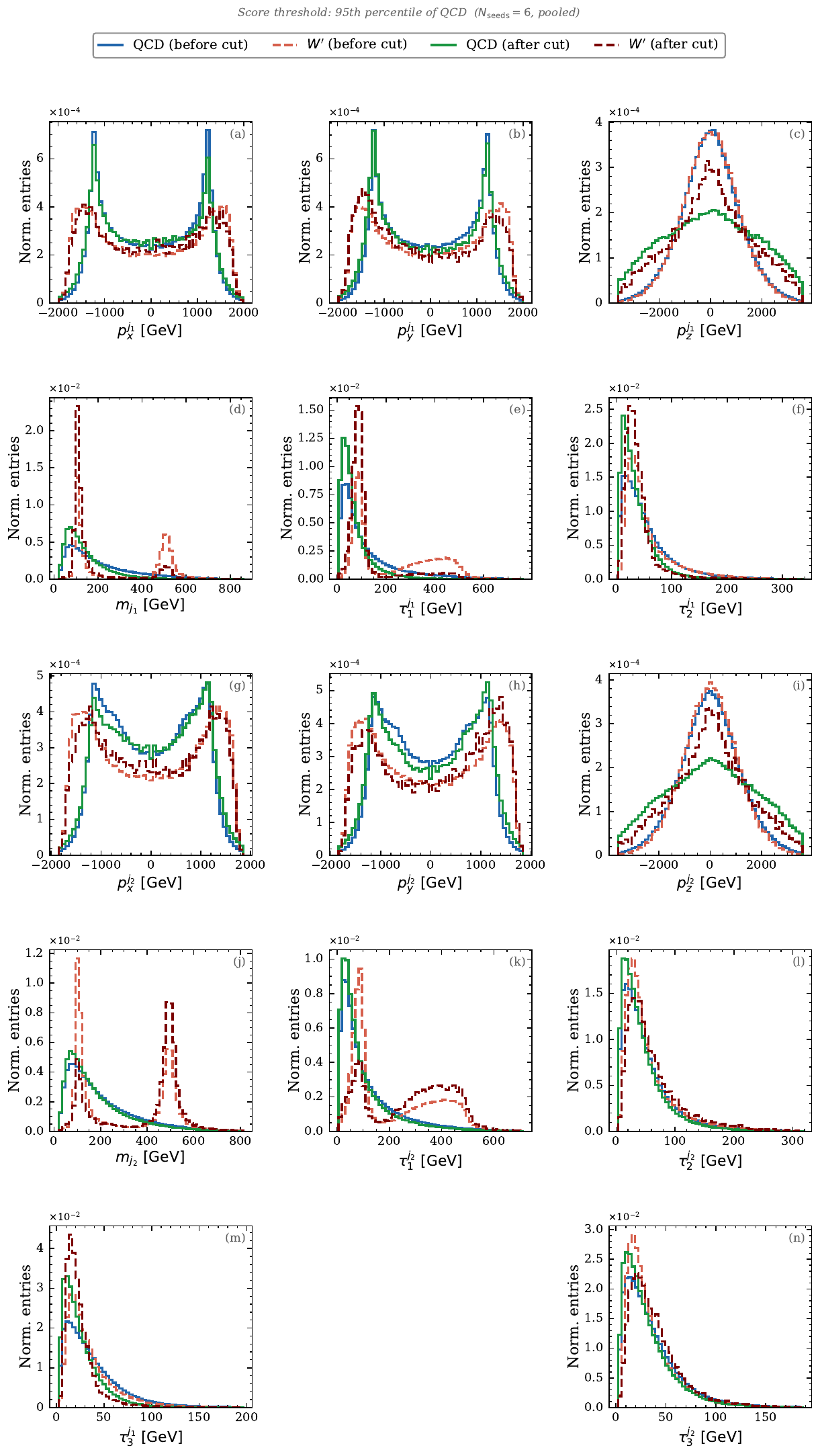}
  \caption{
    Normalized distributions of all 14 input features before (solid) and 
    after (dashed) a selection at a score threshold corresponding to the 
    95th percentile of the QCD background distribution, pooled across six 
    independent random seeds. Blue and green curves show the QCD 
    background before and after the cut; brick-red and crimson curves show 
    the $W^\prime$ signal. The layout follows Fig.~\ref{fig:dataset_overview}.
  }
  \label{fig:feature_shift}
\end{figure}

To further probe the behaviour of the anomaly score and verify that the 
model responds to physically meaningful differences rather than trivial 
kinematic effects, we compare the distributions of all 14 input features 
before and after applying a selection on the score. The threshold is 
defined at a score value corresponding to the 95th percentile of the QCD 
background distribution, corresponding to the operating point used in the 
significance optimization. Figure~\ref{fig:feature_shift} shows, for each 
observable, the QCD background and $W^\prime$ signal both before and after 
the cut, with all distributions pooled across six independent random seeds.

\paragraph{Momentum components.}
The transverse momentum components $p_x$ and $p_y$ for both jets 
(panels~(a),~(b),~(g),~(h)) retain the characteristic bimodal structure 
associated with the back-to-back dijet topology. After the score cut, the 
QCD background becomes slightly narrower with sharper peaks, reflecting a 
preference for more balanced, higher-$p_T$ configurations, while the 
$W^\prime$ signal largely preserves its pre-cut shape with a reduced 
population in the central region between the two lobes. The longitudinal 
momentum $p_z$ (panels~(c) and~(i)) exhibits a visible narrowing for both 
samples after the selection, indicating that the retained events are more 
central in rapidity. The relative shape difference between signal and 
background is nonetheless maintained across the cut, demonstrating that 
the anomaly score is not driven by a simple kinematic boost.

\paragraph{Jet invariant masses.}
The most pronounced feature shift is observed in the jet invariant masses 
$m_{j_1}$ and $m_{j_2}$ (panels~(d) and~(j)). Prior to the cut, the QCD 
background peaks at low mass and falls steeply, while the $W^\prime$ 
signal already displays a secondary structure at higher mass. After the 
selection, the signal develops a strongly enhanced and sharply defined 
peak near $m \approx 80$--$100~\mathrm{GeV}$, consistent with the 
hadronic decay of an on-shell $W$ boson, while the high-mass component is 
significantly suppressed. The QCD background, by contrast, retains its 
smoothly falling spectrum with no sign of artificial sculpting or spurious 
resonant features. This behaviour constitutes direct evidence that the 
physics-aware mass-decorrelation constraint is working as intended: the 
anomaly score enriches genuine signal events while leaving the background 
mass shape intact, a prerequisite for reliable sideband-based bump-hunt 
analyses.

\paragraph{N-subjettiness observables.}
The N-subjettiness observables reveal a coherent and physically motivated 
pattern. For $\tau_1^{j_1}$ and $\tau_1^{j_2}$ (panels~(e) and~(k)), the 
pre-cut signal already exhibits a broad tail at larger values, reflecting 
its characteristic two-prong substructure. After the selection, the signal 
migrates toward lower $\tau_1$ and concentrates around the low-mass peak, 
indicating that the model preferentially retains compact, well-resolved 
hadronic $W$ candidates. The QCD background also shifts slightly toward 
lower values, but the effect is considerably smaller, confirming that the 
score responds primarily to genuine substructure anomalies rather than 
overall jet collimation. A qualitatively similar but weaker trend is seen 
for $\tau_2$ (panels~(f) and~(l)), while the $\tau_3$ distributions 
(panels~(m) and~(n)) show minimal variation across the selection for both 
samples, as expected given that the $W^\prime \to jj$ topology does not 
possess intrinsic three-prong structure. The consistency of these trends 
across both jets and all N-subjettiness orders confirms that the learned 
anomaly criterion is physically coherent and grounded in jet substructure 
rather than in kinematic artifacts.

\subsection{Physics-Motivated Ablation Study}
\label{sec:ablation_strategy}

To quantify the physical role of each architectural component, we perform a systematic ablation study in which individual physics-motivated constraints are removed while all other hyperparameters, training procedures, and random seeds are kept fixed. All comparisons are carried out \emph{seed-by-seed} and subsequently averaged over six independent initializations in order to separate genuine modeling effects from stochastic fluctuations.

The seed-averaged performance differences with respect to the full physics-aware baseline are summarized in Table~\ref{tab:ablation_delta}. Positive shifts in headline metrics such as AUC and $Z_{\mathrm{eff}}$ should not be interpreted as genuine improvements. Instead, they indicate that removing physics constraints allows the model to exploit spurious correlations that artificially enhance raw classification performance while degrading its suitability for resonance searches.

\begin{table}[htbp]
\centering
\caption{Seed-averaged performance differences relative to the baseline model. Positive values
indicate an apparent improvement over the baseline.}
\label{tab:ablation_delta}
\begin{tabular}{lccc}
\hline
Ablation & $\Delta$AUC & $\Delta Z_{\mathrm{eff}}$ & $\Delta \rho(m,\mathrm{score})$ \\
\hline
No mass decorrelation 
& $+0.13 \pm 0.01$ 
& $+0.42 \pm 0.03$ 
& $\mathbf{+0.17 \pm 0.02}$ \\
No KL regularization 
& $+0.12 \pm 0.03$ 
& $+0.25 \pm 0.11$ 
& $+0.03 \pm 0.02$ \\
No diffusion 
& $+0.08 \pm 0.02$ 
& $+0.09 \pm 0.06$ 
& $+0.11 \pm 0.04$ \\
\hline
\end{tabular}
\end{table}

\subsubsection{Impact of Mass Decorrelation}
\label{sec:mass_decorrelation}

\begin{figure}[htbp]
    \centering
    \includegraphics[width=0.65\textwidth]{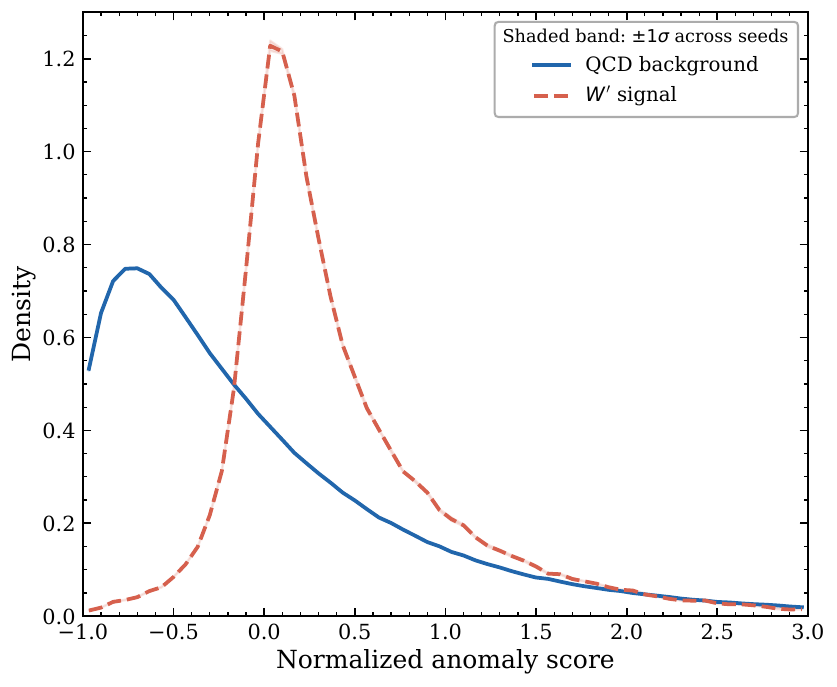}
    \caption{
    Seed-averaged normalized anomaly score distributions for QCD (background) and W$^\prime$ (signal) events. 
    Solid lines show the mean distributions across six seeds, and shaded bands indicate the standard deviation. 
    High-score regions of the non-decorrelated model preferentially select specific mass ranges.
    }
    \label{fig:no_mass_score_dist}
\end{figure}

Among all components, the explicit mass decorrelation constraint has the largest physical impact. Removing it produces a substantial apparent increase in sensitivity, as visible in the SIC curves (Figure~\ref{fig:baseline_sic}) and in the seed-averaged ROC curve (Figure~\ref{fig:baseline_roc}), where the non-decorrelated model systematically outperforms the baseline in terms of raw classification metrics. The anomaly score distributions shown in Figure~\ref{fig:no_mass_score_dist} further confirm a stronger separation between QCD background and W$^\prime$ signal.

This apparent gain is, however, driven by the exploitation of trivial kinematic correlations. The non-decorrelated model preferentially assigns large anomaly scores to events in specific mass regions, resulting in artificial sculpting of the QCD background. Such behavior violates the assumptions underlying sideband-based background estimation and leads to unreliable significance estimates in a realistic search.

To disentangle genuine substructure sensitivity from mass-induced effects,
we apply two standard post-processing decorrelation strategies to the no-mass model:

\begin{itemize}
    \item \textbf{Planing (reweighting)}~\cite{Collins:2018epr}:
    histogram-based reweighting that flattens the mass dependence of the
    background score distribution.

    \item \textbf{DDT (linear decorrelation)}~\cite{Dolen:2016thinking}:
    subtraction of the linear correlation between the anomaly score and the
    reconstructed mass.

    \item \textbf{Planing + DDT}: sequential application of both techniques.
\end{itemize}

\begin{figure}[htbp]
    \centering
    \includegraphics[width=0.75\textwidth]{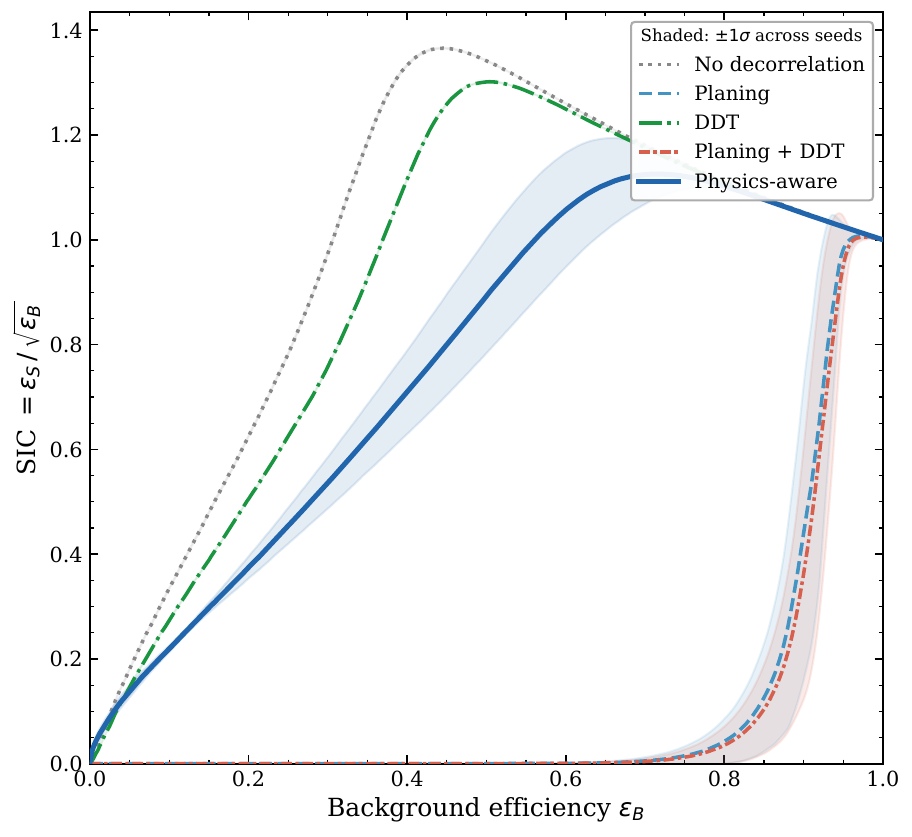}
    \caption{Seed-averaged SIC curves for different mass decorrelation strategies applied to the no-mass model: no decorrelation, Planing, DDT, Planing+DDT, and the baseline physics-aware model. Shaded bands indicate the standard deviation across six random seeds. The Planing and Planing+DDT curves overlap closely, indicating that the additional DDT step does not significantly alter the score distribution when Planing reweighting is already applied. While removing mass decorrelation increases apparent sensitivity, it introduces a substantial correlation with the reconstructed mass, leading to unreliable background estimates in a realistic bump-hunt analysis.}
    \label{fig:mass_decor_sic}
\end{figure}

The resulting SIC curves are compared to the physics-aware baseline in Figure~\ref{fig:mass_decor_sic}. All decorrelation strategies reduce the maximum SIC relative to the raw no-mass model, demonstrating that part of the apparent sensitivity originates from mass correlations. At the same time, they significantly suppress the dependence of the anomaly score on the reconstructed mass, restoring the conditions required for a robust bump hunt.

The statistical stability of these strategies is quantified in Table~\ref{tab:type1_error}, which reports the false-positive rate (FPR) for pure QCD at a fixed 95\% anomaly-score threshold. All decorrelated methods and the physics-aware model maintain stable FPR values close to the nominal 5\%, indicating controlled Type-I error and reliable background behavior in the signal region.

\begin{table}[htbp]
\centering
\begin{tabular}{lcc}
\hline
Method & FPR mean & FPR std \\
\hline
No decorrelation & 0.0509 & 0.00001 \\
Planing & 0.0509 & 0.00001 \\
DDT & 0.0516 & 0.00002 \\
Planing + DDT & 0.0516 & 0.00002 \\
Physics-aware & 0.0518 & 0.00025 \\
\hline
\end{tabular}
\caption{Type-I error (false positive rate) for different mass decorrelation strategies at the 95th percentile anomaly score threshold, averaged over six random seeds. Decorrelated and physics-aware models maintain a safe FPR while mitigating mass sculpting.}
\label{tab:type1_error}
\end{table}

These results demonstrate that explicit mass decorrelation is essential for discovery-oriented anomaly detection. While removing it increases raw classification metrics, it leads to background sculpting and artificially inflated significance. The physics-aware model achieves a lower but physically meaningful sensitivity by ensuring that high anomaly scores are driven by genuine jet substructure deviations rather than kinematic artifacts.

\subsubsection{Role of Bayesian Regularization (KL)}

Removing the KL regularization term results in a modest average increase in headline performance metrics, as reflected by the positive shifts in AUC and $Z_{\mathrm{eff}}$ reported in Table~\ref{tab:ablation_delta}. The seed-averaged ROC curve for the no-KL configuration (Figure~\ref{fig:baseline_roc}) shows a slightly enhanced true-positive rate across most of the background-efficiency range compared to the baseline model.

\begin{figure}[htbp]
    \centering
    \includegraphics[width=0.65\textwidth]{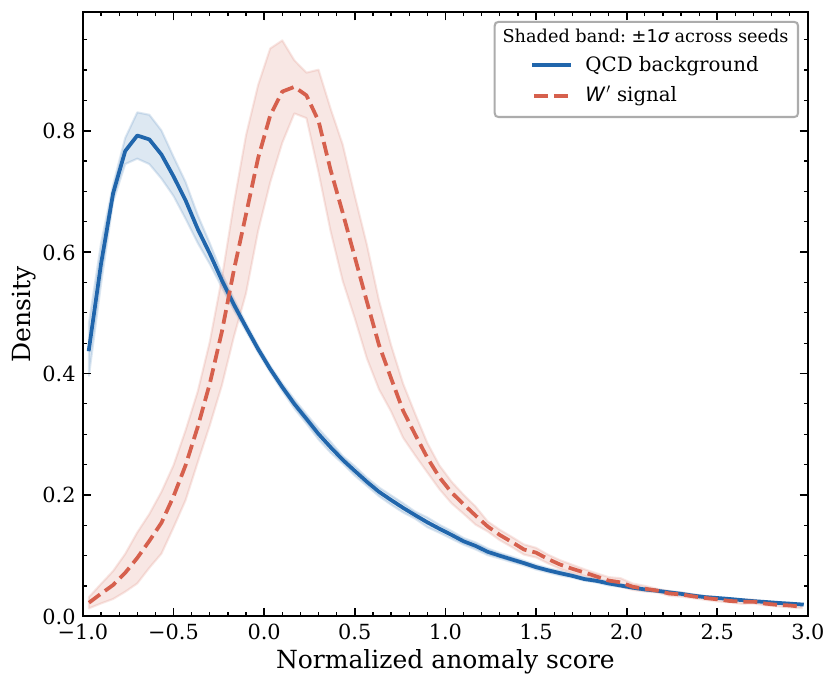}
    \caption{
    Seed-averaged normalized anomaly score distributions for QCD (background) and W$^\prime$ (signal) events. Solid lines show the mean distributions across six seeds, and shaded bands indicate the standard deviation.
    }
    \label{fig:no_kl_score_dist}
\end{figure}

A similar trend is observed in the anomaly-score distributions shown in Figure~\ref{fig:no_kl_score_dist}, where signal events are pushed towards higher scores while maintaining a background-dominated low-score region. This behavior is consistent with a more flexible latent representation that is less constrained by the variational prior.

However, this apparent gain in sensitivity is accompanied by a clear increase in seed-to-seed variability, visible as a broader uncertainty band in both the ROC and SIC curves. The larger spread indicates that the learned latent space becomes more dependent on stochastic optimization effects when the KL term is removed. While the mean mass–score correlation remains consistent with zero, the increased fluctuations across seeds point to reduced robustness of the anomaly ranking.

From a physics perspective, this behavior highlights the role of Bayesian regularization in stabilizing unsupervised anomaly detection. The KL term constrains the latent distribution to remain close to a smooth prior, preventing the model from learning seed-specific features that do not correspond to genuine
substructure differences. This leads to:

\begin{itemize}
    \item improved reproducibility across independent training realizations,
    \item more stable threshold selection for significance optimization,
    \item better calibrated anomaly scores.
\end{itemize}

In the context of resonance searches at the LHC, such stability is essential. Discovery strategies rely on a consistent mapping between anomaly score and background efficiency, and large seed-dependent fluctuations would translate into uncontrolled systematic uncertainties. The KL regularization therefore acts as a physics-motivated constraint that trades a small amount of apparent
sensitivity for significantly improved robustness and interpretability.


\subsubsection{Effect of Latent Diffusion}

Removing the latent diffusion component leads to a clear degradation in robustness, despite only moderate changes in the mean performance metrics. As shown in Figures~\ref{fig:baseline_roc} and~\ref{fig:baseline_sic}, the seed-averaged ROC and SIC curves remain broadly comparable to the baseline in terms of their central values, but exhibit significantly larger seed-to-seed variations.

\begin{figure}[htbp]
    \centering
    \includegraphics[width=0.65\textwidth]{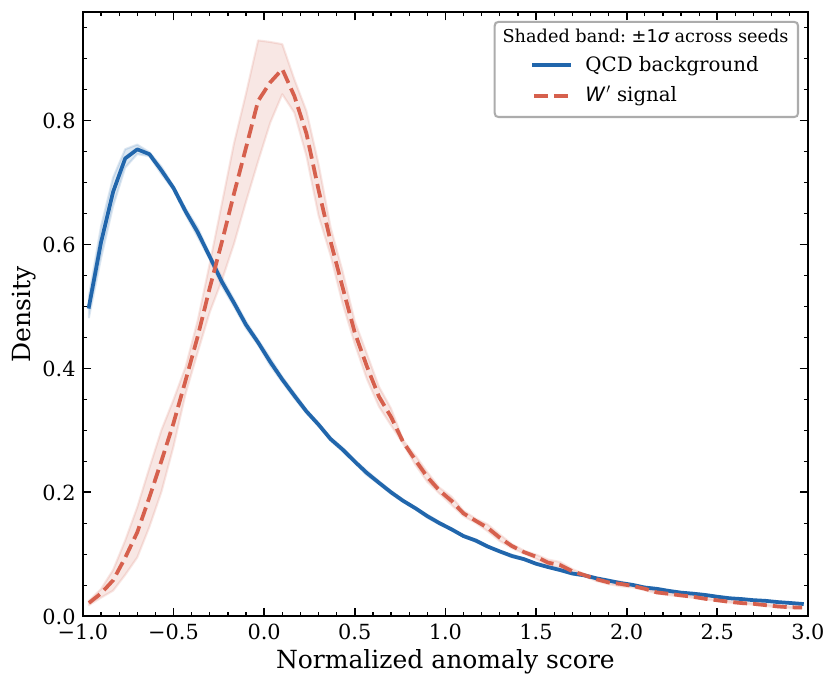}
    \caption{
    Seed-averaged normalized anomaly score distributions for QCD (background) and W$^\prime$ (signal) events. Solid lines show the mean distributions across six seeds, and shaded bands indicate the standard deviation.
    }
    \label{fig:no_diffusion_score_dist}
\end{figure}

This behavior is also visible in the anomaly score distributions (Figure~\ref{fig:no_diffusion_score_dist}), where the separation between QCD and W$^\prime$ events becomes less smooth and more irregular. The resulting score structure indicates a fragmented latent representation, in which nearby events in feature space are not mapped to a coherent anomaly-density ordering.

While the average mass--score correlation remains small, the increased spread across seeds is consistent with the larger standard deviations reported in Table~\ref{tab:seed_stability}. This demonstrates that the absence of diffusion primarily affects the \emph{geometry} of the latent space rather than the mean classification performance.

From a modeling perspective, latent diffusion acts as a stochastic smoothing mechanism that:

\begin{itemize}
    \item enforces continuity of the learned data manifold,
    \item suppresses local overfitting in low-density regions,
    \item stabilizes the anomaly ranking across independent training runs.
\end{itemize}

These properties are particularly important for unsupervised searches, where rare signal-like events populate sparsely sampled regions of phase space. Without diffusion, the anomaly score becomes sensitive to small fluctuations in the training dynamics, leading to reduced reproducibility.

In the context of LHC resonance searches, this translates into a less reliable mapping between anomaly score and background efficiency. The diffusion-driven regularization therefore improves the statistical robustness of the method without introducing mass-dependent features or artificially enhancing the raw sensitivity.

Overall, these results show that latent diffusion is essential for learning a smooth and stable anomaly-density estimator, ensuring that the observed performance reflects genuine substructure differences rather than seed-dependent training artifacts.

\subsubsection{Seed Stability and Robustness}

To quantify the impact of stochastic training effects, all model variants are evaluated across six independent random seeds. The corresponding mean performance and standard deviations are reported in Table~\ref{tab:seed_stability}.

The baseline physics-aware configuration exhibits stable performance across seeds, with moderate variations in AUC and $Z_{\mathrm{eff}}$ and a consistently small mass--score correlation. This behavior indicates that the combined effect of mass decorrelation, Bayesian regularization, and latent diffusion leads to a well-constrained and reproducible anomaly-density estimator.

In contrast, removing individual components alters the stability pattern in a characteristic way. The no-mass-decorrelation model shows artificially small seed-to-seed variations together with a significant positive mass--score correlation. This apparent stability does not reflect a more robust model, but rather the exploitation of a deterministic kinematic feature, which reduces the effective learning complexity while introducing background sculpting.

The removal of the KL regularization term leads to the largest increase in the variance of $Z_{\mathrm{eff}}$, demonstrating that Bayesian regularization plays a central role in controlling fluctuations of the latent representation across independent training realizations. Similarly, the no-diffusion configuration shows increased spread compared to the baseline, confirming that diffusion dynamics contribute to smoothing the learned data manifold and stabilizing the anomaly ranking.

These results show that seed stability provides a complementary diagnostic to headline performance metrics. A higher AUC obtained at the cost of increased mass correlation or reduced reproducibility does not correspond to improved physics sensitivity. Instead, the baseline model achieves the most reliable performance in the sense required for data-driven background estimation and statistical inference.

\begin{table}[htbp]
\centering
\caption{Mean performance across random seeds. Quoted uncertainties correspond to the standard
deviation over six independent training runs.}
\label{tab:seed_stability}
\begin{tabular}{lccc}
\hline
Model Variant & AUC & $Z_{\mathrm{eff}}$ & $\rho(m,\mathrm{score})$ \\
\hline
Baseline (full) 
& $0.59 \pm 0.03$ 
& $2.27 \pm 0.07$ 
& $-0.10 \pm 0.03$ \\
No mass decorrelation 
& $0.72 \pm 0.01$ 
& $2.69 \pm 0.01$ 
& $+0.07 \pm 0.01$ \\
No KL regularization 
& $0.71 \pm 0.02$ 
& $2.52 \pm 0.10$ 
& $-0.07 \pm 0.02$ \\
No diffusion 
& $0.67 \pm 0.01$ 
& $2.36 \pm 0.05$ 
& $+0.01 \pm 0.03$ \\
\hline
\end{tabular}
\end{table}

Additional robustness checks with modified network capacity, including wider architectures, are reported in Appendix~\ref{app:arch_checks}.

\subsection{Summary of Physics Ablation Results}

The ablation study demonstrates that the individual components of the training objective control complementary aspects of the anomaly-detection problem.

Mass decorrelation is essential for ensuring that the anomaly score is not driven by trivial kinematic features, thereby preserving the validity of sideband-based background estimation. Bayesian regularization primarily stabilizes the latent representation across independent training runs, leading to reliable uncertainty estimates and consistent discovery thresholds. Latent diffusion enforces a smooth data manifold and improves the reproducibility of the anomaly ranking, which is crucial for rare-event searches in sparsely populated regions of phase space.

While several ablations show higher AUC and $Z_{\mathrm{eff}}$, these apparent improvements are associated with increased mass correlations or reduced seed stability and therefore do not correspond to genuine gains in physics sensitivity. The full physics-aware configuration provides the most robust and interpretable performance, even though it is not explicitly optimized for maximal discovery significance.

This behavior highlights a central principle of anomaly detection for LHC searches: physically consistent modeling of the background leads to reliable sensitivity, whereas aggressive optimization of classification metrics alone can produce misleading results.

\subsection{Model-Architecture-Motivated Ablation Study}

\subsubsection{Latent Representational Capacity}

\begin{figure}[htbp]
    \centering
    \includegraphics[width=0.48\textwidth]{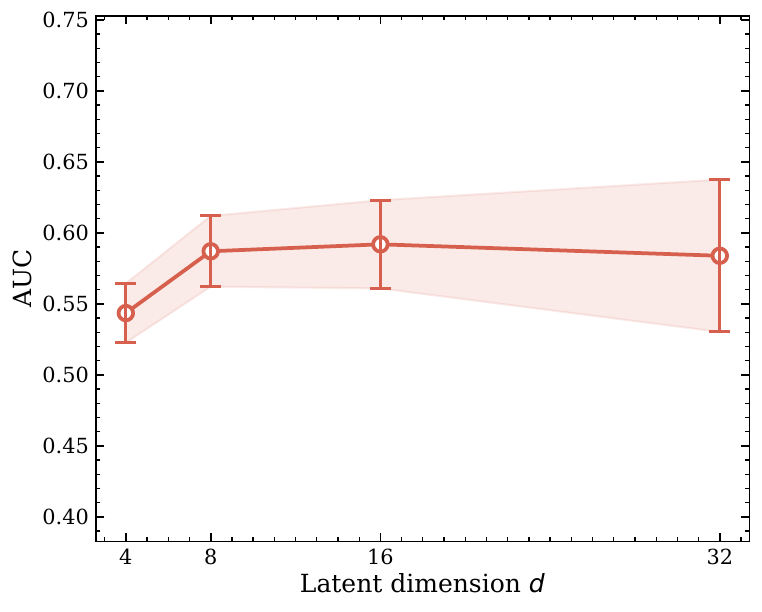}
    \includegraphics[width=0.48\textwidth]{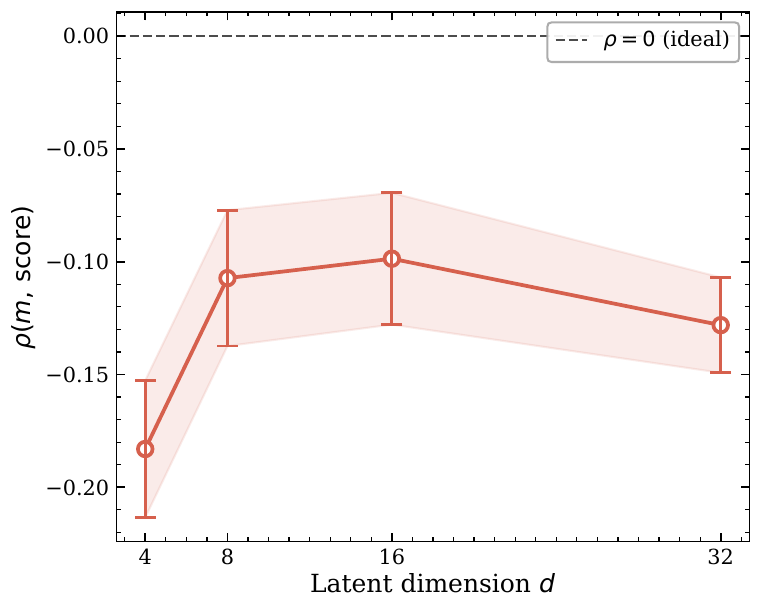}
    \caption{
    Seed-averaged performance as a function of latent dimensionality.
    \textbf{Left:} Mean AUC with standard deviation across six random seeds.
    \textbf{Right:} Pearson correlation between anomaly score and invariant mass.
    The optimal trade-off between expressivity and stability is achieved at
    $d=16$, while mass decorrelation remains effective across all dimensions.
    }
    \label{fig:latent_dim_ablation}
\end{figure}

To investigate the role of latent-space dimensionality, we evaluate the baseline architecture using latent dimensions $d = 4, 8, 16,$ and $32$, while keeping all other training settings fixed. This study probes the balance between information compression, expressivity, and regularization in the uncertainty-aware latent representation.

Figure~\ref{fig:latent_dim_ablation} shows that at low dimensionality ($d=4$), the model exhibits degraded discrimination power and increased seed-to-seed variability, indicating insufficient capacity to encode relevant jet substructure information. Increasing the latent dimension improves performance, with $d=16$ providing the most stable trade-off between expressivity and regularization. Further increasing the latent dimension to $d=32$ yields only marginal gains in average AUC while introducing larger statistical fluctuations across seeds, consistent with mild over-capacity relative to the available training statistics.

The seed-averaged Pearson correlation coefficient $\rho(m,\mathrm{score})$ provides a quantitative measure of mass sculpting induced by the anomaly score, with values consistent with zero indicating that high-score selections do not bias the invariant-mass distribution.

In practical analyses, this ensures that selecting high-score events does not distort background mass sidebands used for data-driven background estimation.

Importantly, the mass--score correlation remains consistently suppressed across all latent dimensions. This demonstrates that the physics-aware constraints enforce robust mass decorrelation independently of representational capacity, indicating that the model’s physical behavior is not fine-tuned to a specific latent dimension.

\subsubsection{Diffusion Process Sensitivity}

\begin{figure}[htbp]
    \centering
    \includegraphics[width=0.48\textwidth]{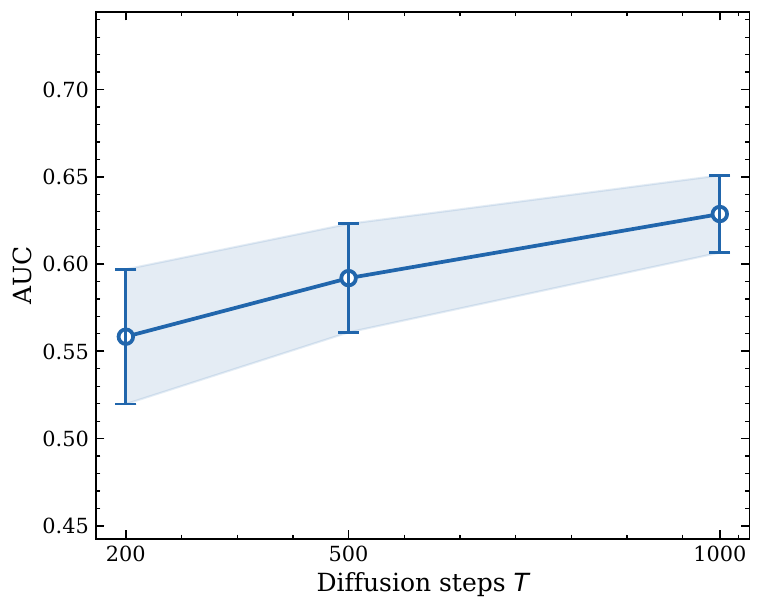}
    \includegraphics[width=0.48\textwidth]{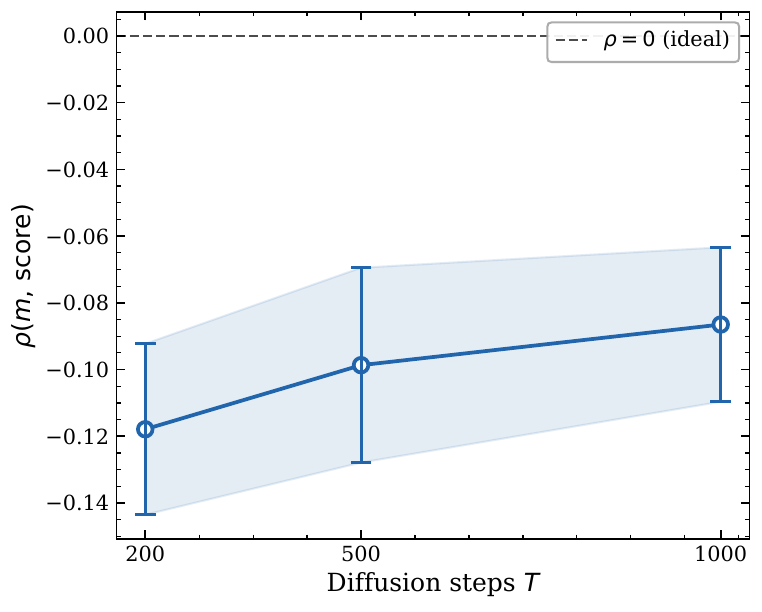}
    \caption{
    Seed-averaged performance as a function of diffusion depth.
    \textbf{Left:} Mean AUC with standard deviation across seeds.
    \textbf{Right:} Pearson correlation between anomaly score and invariant mass.
    Increasing diffusion depth improves stability and smoothness of the latent
    representation without introducing mass-dependent biases.
    }
    \label{fig:diffusion_ablation}
\end{figure}

We study the sensitivity of the framework to the depth of the latent diffusion process by varying the number of diffusion steps $T = 200, 500,$ and $1000$, while keeping all other settings fixed. This analysis assesses how the granularity of the noising–denoising dynamics influences the learned latent manifold and downstream anomaly scores.

As shown in Figure~\ref{fig:diffusion_ablation}, shallow diffusion horizons ($T=200$) lead to increased seed-to-seed variability, indicating incomplete regularization of the latent space.
Increasing the diffusion depth to $T=500$ significantly improves stability and
reproducibility, while further increasing to $T=1000$ yields only marginal additional gains without observable degradation in training behavior.

Across all diffusion depths, the mass--score correlation remains close to zero, confirming that diffusion acts as a generative regularizer rather than exploiting kinematic information. These results demonstrate that latent diffusion enhances robustness and smooth latent geometry without requiring fine-tuned diffusion schedules.

\subsection{Summary of Model-Architecture Ablation Results}

The architecture-motivated ablation studies demonstrate that the proposed framework exhibits stable performance across a broad range of representational and generative configurations.
While latent dimensionality and diffusion depth control the balance between expressivity and regularization, they do not qualitatively alter the physical behavior of the anomaly score or reintroduce mass sculpting.

Crucially, the baseline configuration lies within a wide plateau of stable and reproducible performance rather than at a finely tuned optimum. This indicates that the observed anomaly-detection capabilities arise from the physics-aware design of the framework, supporting its robustness and scalability
for future collider applications.

\section{Conclusion}

We have presented a physics-informed Bayesian latent diffusion framework for fully unsupervised anomaly detection in collider data. By combining uncertainty-aware Bayesian encoding, latent-space diffusion modeling, and explicit physics-motivated regularization, the proposed approach prioritizes robustness, interpretability, and physical consistency over isolated gains in headline performance. Through seed-averaged evaluations and systematic ablation studies, we demonstrated that mass decorrelation is essential for suppressing kinematic sculpting, Bayesian regularization stabilizes training across random initializations, and latent diffusion enforces a smooth and robust background manifold. While improvements in peak anomaly-detection metrics are moderate, the framework exhibits stable behavior across seeds, suppressed mass–score correlations, and reliable uncertainty control. These properties position the method as a reliable and physics-aware foundation for discovery-oriented anomaly searches at the LHC, where control of systematic effects is as critical as raw sensitivity.
\section{Future Work}

Future work will extend this framework to more realistic and information-rich jet representations, including the RODEM Jet Datasets~\cite{Zoch:2024eyp}, which provide both jet-level observables and detailed constituent information. We plan to develop transformer-based encoders capable of processing variable-length constituent inputs and embedding jets into structured latent representations, coupled to latent-space diffusion models that learn the Standard-Model background manifold from data alone. Physics-aware reconstruction constraints will enforce consistency between high-level observables and constituent kinematics. This extension will enable fully unsupervised, model-agnostic searches for new-physics signatures directly at the detector level, strengthening the connection between modern generative modeling and practical collider discovery strategies.
\section*{Acknowledgments}

We gratefully acknowledge the support of the INFN Padova computing facility, whose high-performance computing resources were essential for the training and evaluation of the models presented in this work. During the initial stages of the project, part of the model development and exploratory studies was performed using the free computational resources provided by Google Colab.

\bibliographystyle{unsrt}
\bibliography{references}

\section{Appendix}
\label{app:additional_results}

This appendix collects supplementary results and diagnostic studies supporting the main analysis. These materials are provided to assess training stability, robustness to stochastic optimization, internal behavior of the anomaly score, and sensitivity to architectural hyperparameters. Unless otherwise stated, all quantities shown here correspond to seed-averaged results across six independent random seeds.

\subsection{Training Hyperparameter Variation}
\label{app:training_loss_all_components}

\begin{figure}[htbp]
    \centering
    \begin{subfigure}[b]{0.50\textwidth}
        \includegraphics[width=\textwidth]{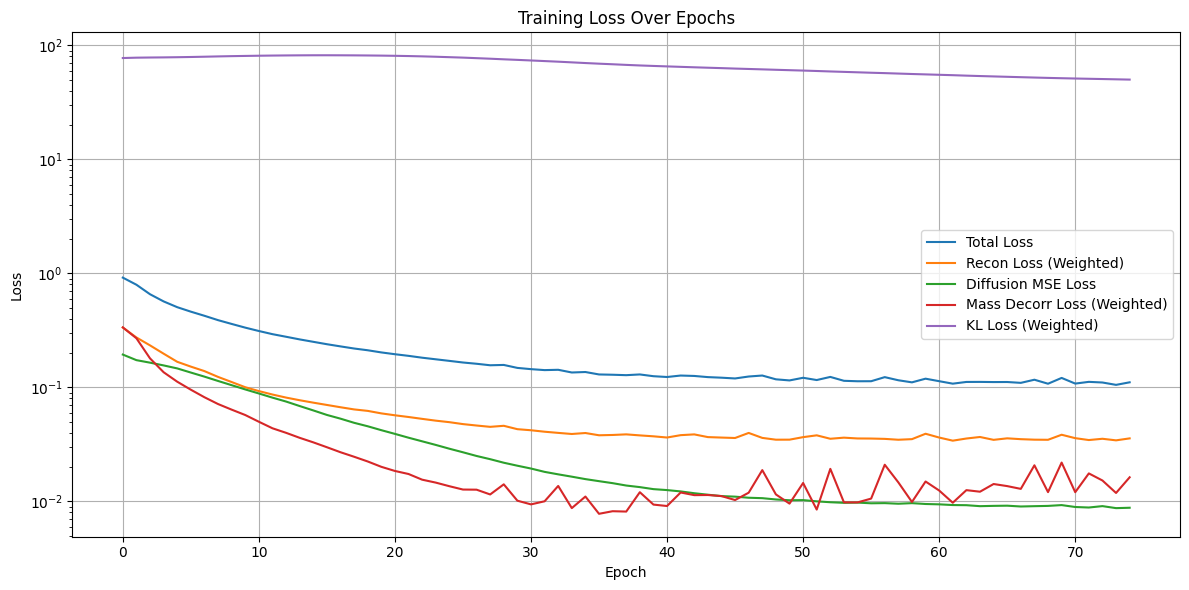}
        \caption{Run 1}
    \end{subfigure}
    \hfill
    \begin{subfigure}[b]{0.50\textwidth}
        \includegraphics[width=\textwidth]{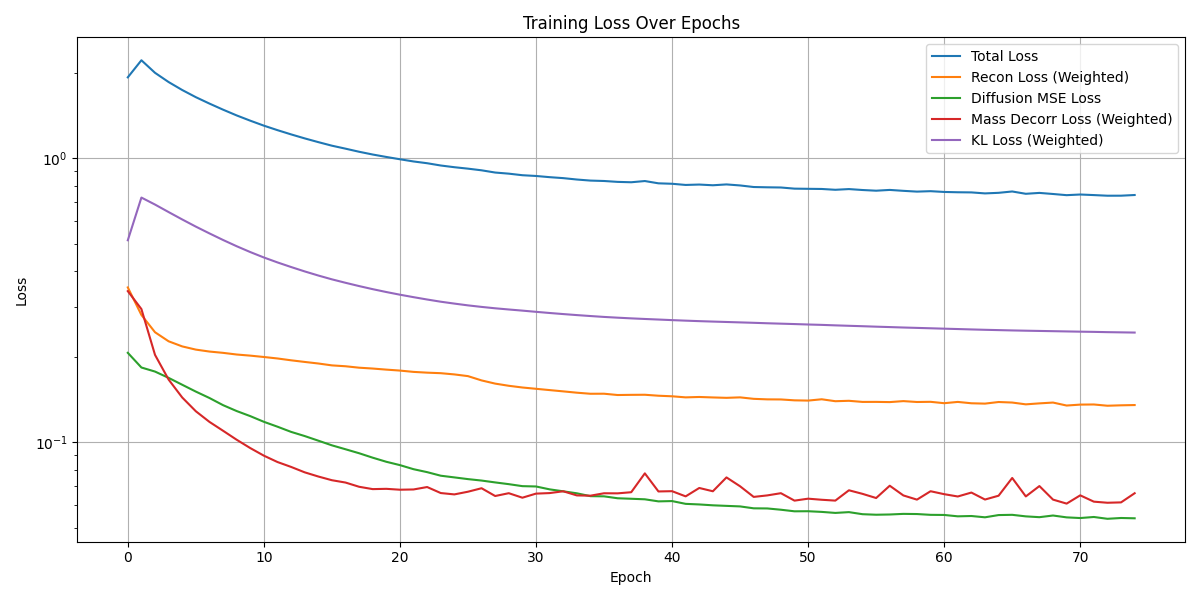}
        \caption{Run 2}
    \end{subfigure}
    \hfill
    \begin{subfigure}[b]{0.50\textwidth}
        \includegraphics[width=\textwidth]{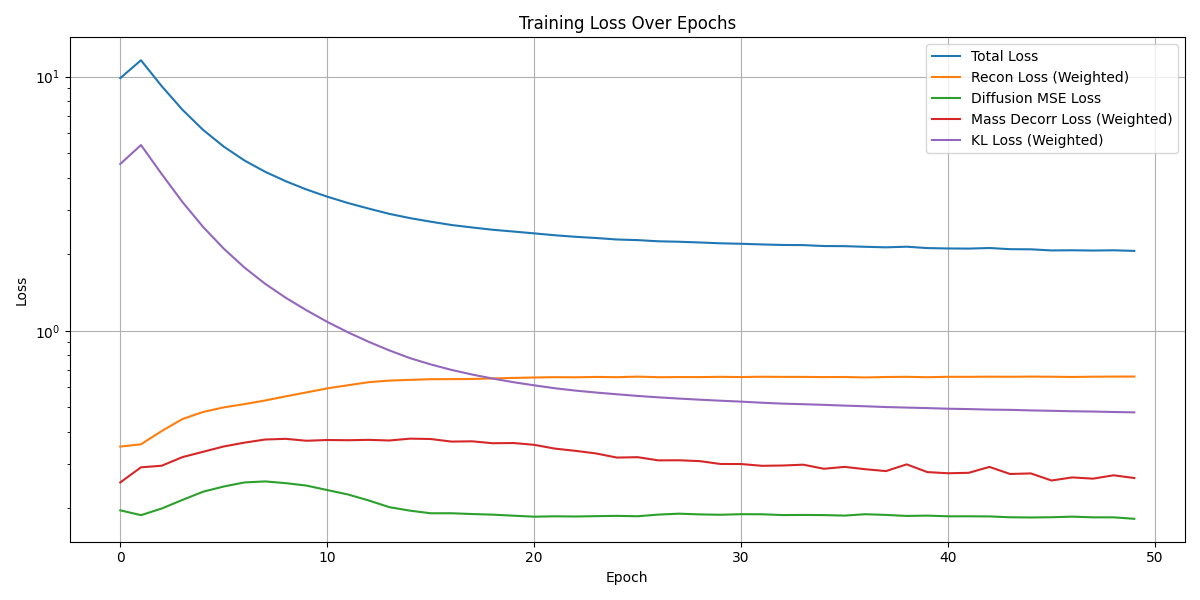}
        \caption{Run 3}
    \end{subfigure}
    
    \caption{
    Training loss curves for three independent runs with varied hyperparameter configurations (exact values not recorded). 
    Each plot shows all loss components: KL divergence, mass-decorrelation loss, diffusion loss, reconstruction loss, and total loss. 
    }
    \label{fig:training_loss_all_components}
\end{figure}

\subsection{Training Diagnostics for Ablation Studies}
\label{app:training_diagnostics}

\begin{figure}[htbp]
    \centering

    \begin{subfigure}{0.48\textwidth}
        \centering
        \includegraphics[width=\textwidth]{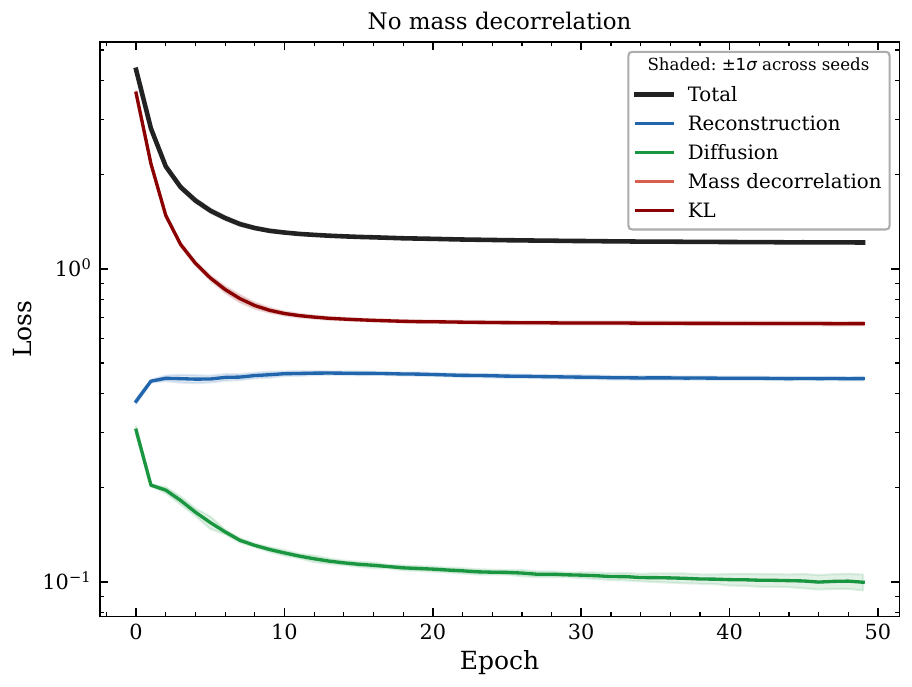}
        \caption{No mass decorrelation}
        \label{fig:loss_no_mass}
    \end{subfigure}
    \hfill
    \begin{subfigure}{0.48\textwidth}
        \centering
        \includegraphics[width=\textwidth]{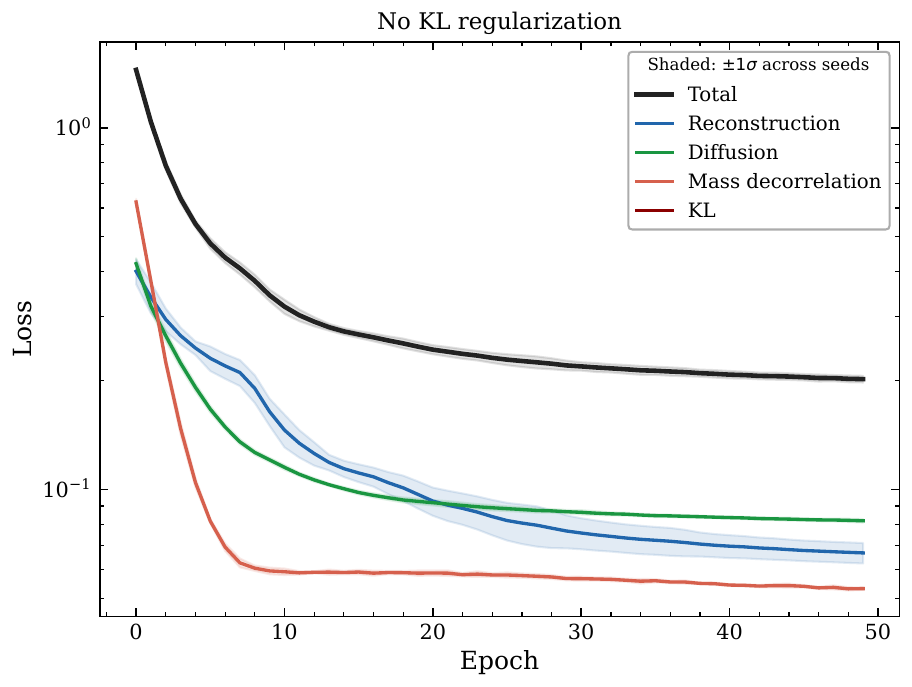}
        \caption{No KL regularization}
        \label{fig:loss_no_kl}
    \end{subfigure}

    \vspace{0.4cm}

    \begin{subfigure}{0.48\textwidth}
        \centering
        \includegraphics[width=\textwidth]{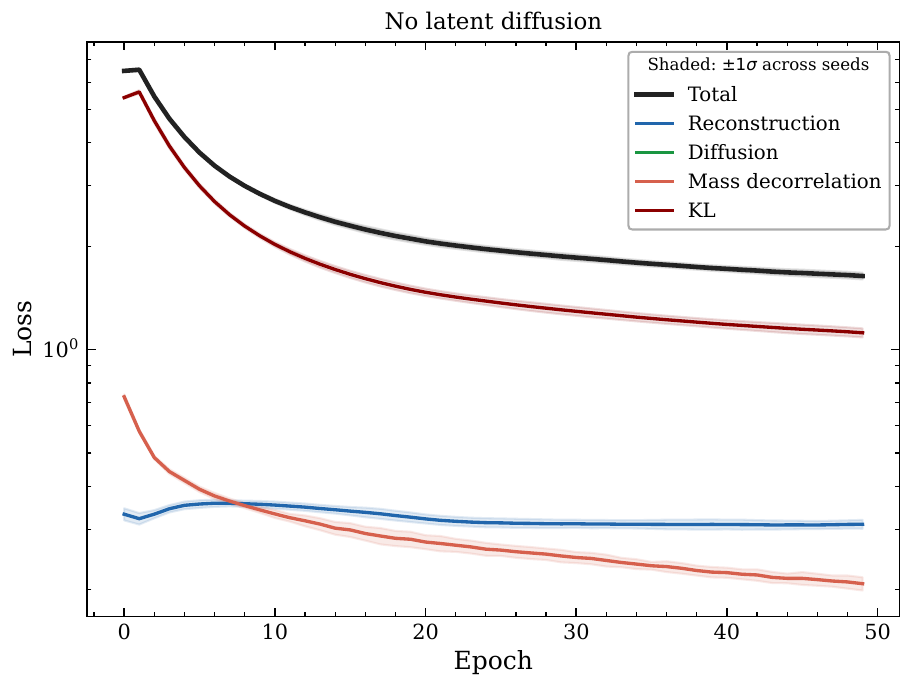}
        \caption{No latent diffusion}
        \label{fig:loss_no_diff}
    \end{subfigure}

    \caption{
    Seed-averaged evolution of training losses for physics-motivated ablation models.
    Solid curves indicate the mean across six independent random seeds, with shaded bands
    representing one standard deviation.
    Deviations from the baseline reflect altered optimization dynamics induced by the
    removal of specific regularization components.
    }
    \label{fig:ablation_training_losses}
\end{figure}

\begin{figure}[htbp]
    \centering

    \begin{subfigure}{0.48\textwidth}
        \centering
        \includegraphics[width=\textwidth]{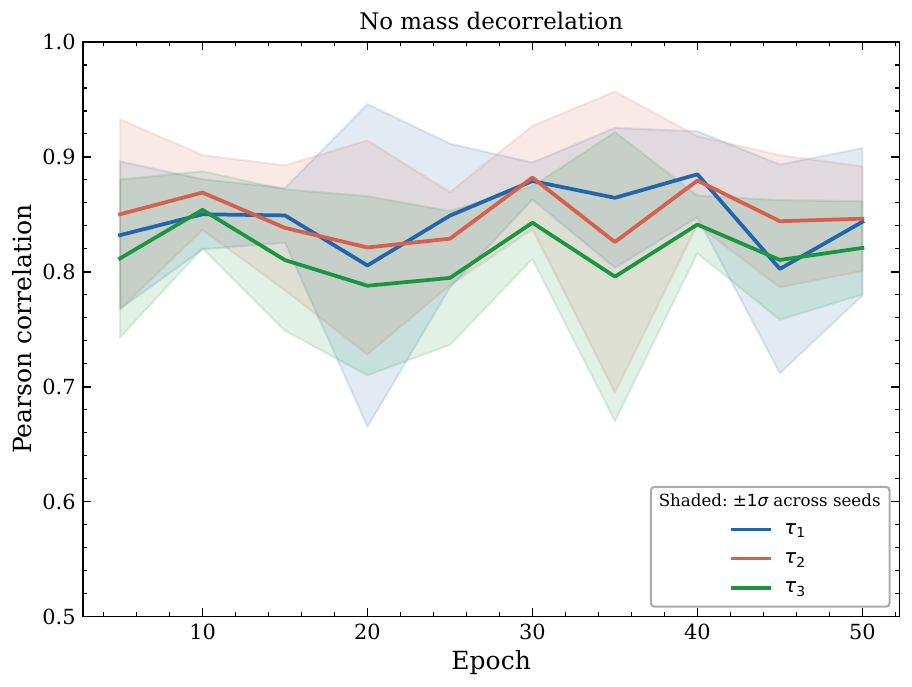}
        \caption{No mass decorrelation}
        \label{fig:tau_no_mass}
    \end{subfigure}
    \hfill
    \begin{subfigure}{0.48\textwidth}
        \centering
        \includegraphics[width=\textwidth]{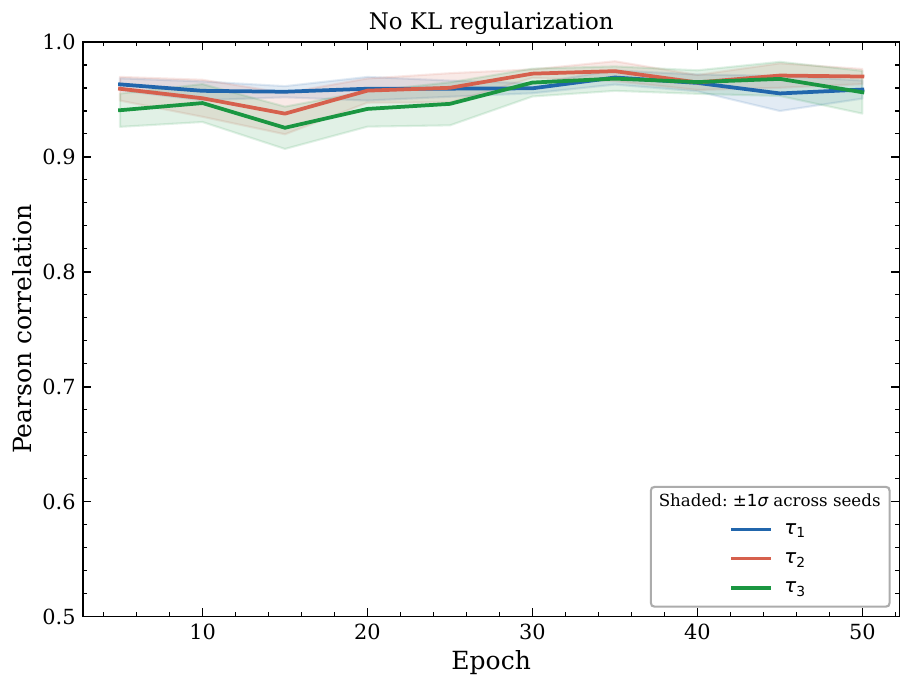}
        \caption{No KL regularization}
        \label{fig:tau_no_kl}
    \end{subfigure}

    \vspace{0.4cm}

    \begin{subfigure}{0.48\textwidth}
        \centering
        \includegraphics[width=\textwidth]{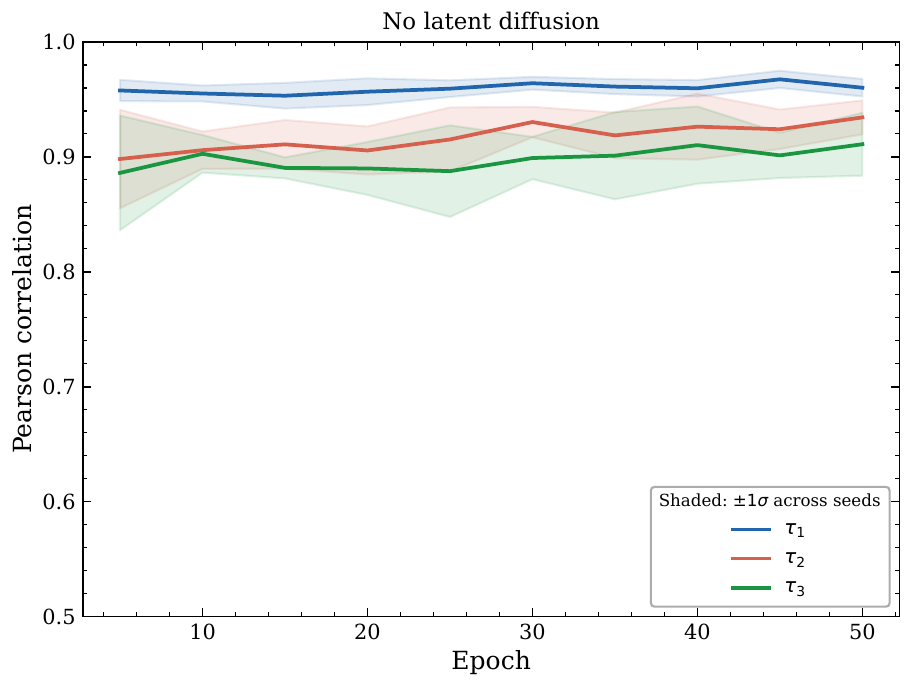}
        \caption{No latent diffusion}
        \label{fig:tau_no_diff}
    \end{subfigure}

    \caption{
    Seed-averaged Pearson correlation coefficients between the anomaly score and jet
    substructure observables $\tau_1$, $\tau_2$, and $\tau_3$.
    Shaded bands indicate seed-to-seed variability.
    The modified convergence patterns illustrate the stabilizing role of physics-aware
    regularization terms.
    }
    \label{fig:ablation_tau_correlations}
\end{figure}

Figures~\ref{fig:ablation_training_losses} and~\ref{fig:ablation_tau_correlations} show 
the seed-averaged training loss curves and jet substructure correlation coefficients for 
the three physics-motivated ablation configurations discussed in Section~\ref{sec:results}. 
These diagnostics are provided for completeness and to support reproducibility of the 
reported results.

\subsection{Per-seed Performance Results}
\label{app:per_seed}

All per-seed results are reported for transparency and reproducibility.
The physics interpretation in the main text is based exclusively on seed-averaged quantities.

\begin{table}[htbp]
\centering
\caption{Per-seed performance for the baseline model with full physics constraints.}
\label{tab:baseline_per_seed}
\begin{tabular}{cccc}
\hline
Seed & AUC & $Z_{\mathrm{eff}}$ & $\rho(m,\mathrm{score})$ \\
\hline
7    & 0.622 & 2.33 & $-0.121$ \\
42   & 0.594 & 2.27 & $-0.069$ \\
123  & 0.533 & 2.16 & $-0.137$ \\
555  & 0.580 & 2.26 & $-0.123$ \\
1251 & 0.595 & 2.25 & $-0.063$ \\
2024 & 0.627 & 2.35 & $-0.079$ \\
\hline
\end{tabular}
\end{table}

\begin{table}[htbp]
\centering
\caption{Per-seed performance without mass decorrelation.}
\label{tab:no_mass_per_seed}
\begin{tabular}{cccc}
\hline
Seed & AUC & $Z_{\mathrm{eff}}$ & $\rho(m,\mathrm{score})$ \\
\hline
7    & 0.717 & 2.69 & $+0.069$ \\
42   & 0.717 & 2.69 & $+0.069$ \\
123  & 0.717 & 2.69 & $+0.067$ \\
555  & 0.716 & 2.69 & $+0.066$ \\
1251 & 0.717 & 2.69 & $+0.069$ \\
2024 & 0.717 & 2.69 & $+0.070$ \\
\hline
\end{tabular}
\end{table}

\begin{table}[htbp]
\centering
\caption{Per-seed performance without KL regularization.}
\label{tab:no_kl_per_seed}
\begin{tabular}{cccc}
\hline
Seed & AUC & $Z_{\mathrm{eff}}$ & $\rho(m,\mathrm{score})$ \\
\hline
7    & 0.687 & 2.43 & $-0.092$ \\
42   & 0.694 & 2.41 & $-0.072$ \\
123  & 0.711 & 2.51 & $-0.080$ \\
555  & 0.717 & 2.56 & $-0.040$ \\
1251 & 0.689 & 2.51 & $-0.100$ \\
2024 & 0.739 & 2.67 & $-0.034$ \\
\hline
\end{tabular}
\end{table}

\begin{table}[htbp]
\centering
\caption{Per-seed performance without latent diffusion.}
\label{tab:no_diffusion_per_seed}
\begin{tabular}{cccc}
\hline
Seed & AUC & $Z_{\mathrm{eff}}$ & $\rho(m,\mathrm{score})$ \\
\hline
7    & 0.668 & 2.27 & $+0.020$ \\
42   & 0.677 & 2.43 & $+0.001$ \\
123  & 0.671 & 2.36 & $-0.039$ \\
555  & 0.670 & 2.34 & $+0.004$ \\
1251 & 0.673 & 2.37 & $+0.044$ \\
2024 & 0.680 & 2.41 & $+0.039$ \\
\hline
\end{tabular}
\end{table}


\subsection{Architecture Robustness Checks}
\label{app:arch_checks}

\begin{figure}[htbp]
    \centering
    \begin{subfigure}{0.48\textwidth}
        \centering
        \includegraphics[width=\textwidth]{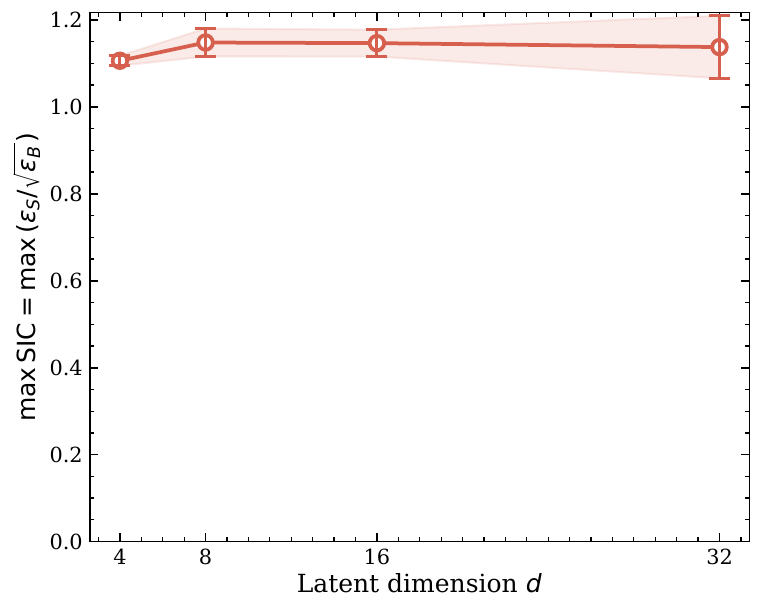}
        \caption{$\max\,\mathrm{SIC}$ vs.\ latent dimension}
    \end{subfigure}
    \hfill
    \begin{subfigure}{0.48\textwidth}
        \centering
        \includegraphics[width=\textwidth]{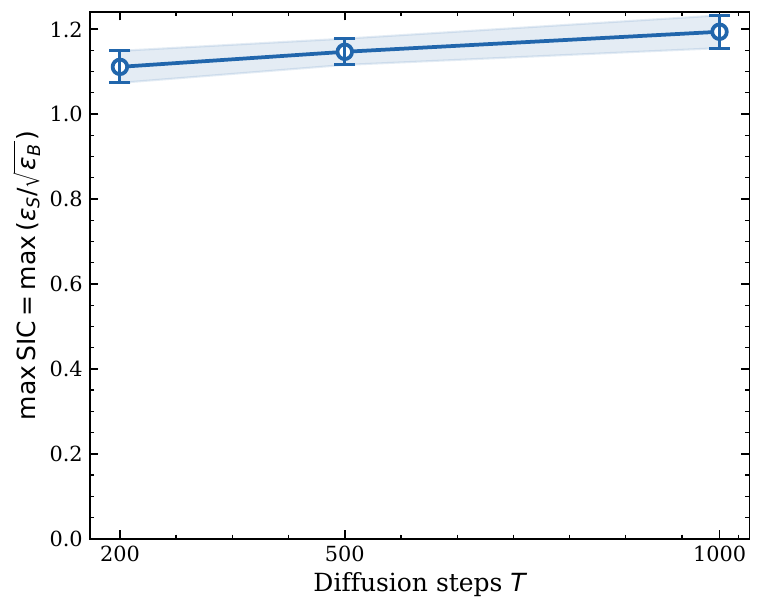}
        \caption{$\max\,\mathrm{SIC}$ vs.\ diffusion depth}
    \end{subfigure}
    \caption{
    Seed-averaged maximum Signal Improvement Characteristic
    $(\max\,\varepsilon_S/\sqrt{\varepsilon_B})$ as a function of
    latent-space dimensionality and diffusion depth.
    Error bars indicate one standard deviation across random seeds.
    The weak dependence confirms that discovery reach is not driven by architectural
    hyperparameter tuning.
    }
    \label{fig:zeff_architecture_check}
\end{figure}

\end{document}